\documentclass[11pt,prd,superscriptaddress,nofootinbib]{revtex4-2}
\usepackage{graphicx}
\usepackage{float}
\usepackage{placeins}
\usepackage[dvipsnames]{xcolor}

\usepackage[colorlinks=true, linkcolor=blue, citecolor=blue, urlcolor=blue]{hyperref}

\begin{document}

\title{Pion Phenomenology from the Thermal Soft-Wall Model of Holographic QCD}

\author{Narmin Nasibova}
\email{n.nesibli88@gmail.com}
\affiliation{Institute of Physics, Ministry of Science and Education,\\  H. Javid Avenue 131, AZ1073, Baku, Azerbaijan}
\affiliation{Center for Theoretical Physics, Khazar University, 41 Mehseti Str.,AZ1096, Baku, Azerbaijan}
\affiliation{The French-Azerbaijani University under Azerbaijan State Oil and Industry University,\\
Nizami str. 183, Baku, Azerbaijan}

\author{Xerxes D. Arsiwalla}
\email{x.d.arsiwalla@gmail.com}
\affiliation{Wolfram Institute for Computational Foundations of Science,\\ Champaign, IL, USA}

\begin{abstract}
\end{abstract}

\maketitle
\thispagestyle{empty}

\vspace{2cm}
\centerline{\bf Abstract} 
\vskip 4mm
Within the framework of the thermal soft-wall model of AdS/QCD, we investigate phenomenological properties of pions at finite temperature. This includes the  electromagnetic (EM) form factor $F_{\pi}(Q^{2}, T)$, the thermal mass $M_{\pi}(T)$, charge radius $r_{\pi}(T)$,  the generalized parton distribution (GPD) $H_{\pi}(x, Q^{2}, T)$, the  charge density $\rho_{\pi}(b, T)$ of the pion, the pion-nucleon coupling constant $g_{\pi NN}(T)$, and  pion-$\Delta$ baryon coupling constant $g_{\pi \Delta \Delta}(T)$ coupling constant at finite temperature. The thermal pion form factor is extrapolated from the zero-temperature case. Subsequently, the GPD is obtained at finite temperature from this form factor. The above-mentioned quantities were analyzed using a thermal dilaton field in the five-dimensional AdS action. Moreover, we determine the theoretical expression for the temperature-dependent pion-nucleon coupling constant, and the pion-$\Delta$ baryon coupling constant using thermal profile functions of the nucleon, $\Delta$ baryon and the pion. Our results show that the values of these quantities decrease with increasing temperature and vanish near the critical temperature $T_c$; except for the pion radius, which diverges at $T_c$. Our results reproduce expected features of low-energy hadron dynamics, thus validating the phenomenological utility of the thermal soft-wall model.


\clearpage
\tableofcontents

\vspace{1cm}
\section{Introduction}

The investigation of hadron physics at finite temperatures provides invaluable insight into the phenomenology of baryons and mesons, as well as the fundamental nature of strong interactions described by Quantum Chromodynamics (QCD).  To study the internal structure and dynamics of hadrons, various theoretical models have been developed over the years. These models are designed to describe hadronic behavior under different physical conditions and energy scales. 

In this context, a particularly relevant class of models based on the AdS/CFT correspondence \cite{Maldacena, Makoto, Anastasia1, Anastasia2} are holographic QCD (or AdS/QCD) models \cite{Boschi1, Boschi2, Erlich, Rold1, Rold2, Karch, Brodsky2012, Branz, Maru1}. Many problems related to the strongly coupled regime of QCD \cite{Paula, Rinaldi} have been addressed within the framework of holographic models. Holographic QCD models have the advantage that they incorporate both the small and the large regime of momentum transfer, that is, they do not impose a limit on $Q^2$. In contrast, experiments and alternative models such as QCD Sum Rules \cite{tahm, tahm1, tahm2, Azizi1} typically focus on a limited momentum transfer window. Furthermore, within the realm of non-perturbative QCD, the AdS/QCD framework has proven to be a powerful tool for investigating properties of hadrons that cannot be investigated via perturbative methods. In particular, the soft-wall model of AdS/QCD provides an effective approach to modeling the dynamics of hadrons, including their internal structure and interactions with other fields, in both, the vacuum, as well as within a hot medium. A key feature  of the AdS/QCD soft-wall model is that it produces linear Regge trajectories in the mass spectrum of mesons \cite{Karch, Paula, Rinaldi}. This model incorporates the effects of confinement and mass generation in QCD by introducing a background dilaton field that breaks the  conformal symmetry of the AdS background. Consequently, this leads to a more realistic description of hadrons, allowing for the computation of key observables, such as the hadron form factor, parton distribution function (PDF), GPD  \cite{Gutsche30, Hashamipour, Hashamipour1, Hashamipour2, Puhan2025}, hadron masses, hadron radii, charge density, and coupling constants of hadrons with other hadrons in a manner that quantitatively captures the low-energy dynamics of QCD.

In fact, the above AdS/QCD framework has been extended to include finite-temperature effects: what is referred to as the AdS/QCD soft-wall model with a temperature-dependent dilaton field (the thermal soft-wall model) \cite{Gutsche2, Gutsche3}. This allows one to investigate the evolution of QCD observables under thermal conditions. Refs.~\cite{Gutsche2, Gutsche3} have elucidated the finite-temperature theory of mesons and baryons within the AdS/QCD soft-wall model. In Refs.~\cite{Gutsche2, Gutsche3, Nasibova1, NasibovaL, NasibovaF}, the authors have obtained expressions for profile functions of mesons and baryons, the meson-baryon form factor, and the coupling constant at finite temperature by applying thermal profile functions in their analysis. On the dual AdS gravity side, the critical temperature corresponds to the temperature of a first-order Hawking-Page phase transition in the AdS spacetime with a black hole, while it is the confinement-deconfinement phase transition temperature in the gauge theory (on the boundary). This critical temperature has been estimated at $T_{c} = 0.122$ $GeV$ in the hard-wall model, $T_{c}=0.191$ $GeV$ in the soft-wall model of holographic QCD (see \cite{Gutsche2}), and $T_{c}=0.192\pm0.007$ $GeV$ in lattice QCD \cite {Cheng5}.  

The pion as a light  pseudoscalar meson plays a fundamental role in low-energy Quantum Chromodynamics, emerging as a pseudo-Goldstone boson due to spontaneous chiral symmetry breaking \cite{Ayala, Gherghetta}. Its well-measured mass and decay constants make it an ideal candidate for testing theoretical frameworks such as the AdS/QCD soft-wall model and comparing it with experimental data. The pion is also an ideal candidate for investigating the mechanism of chiral symmetry breaking.


One of the most intriguing aspects of hadron physics is the thermal behavior of composite particles such as the pion, which plays a central role in low-energy hadron physics (below the confinement-deconfinement phase transition temperature) of QCD. The pion EM form factor encodes information about the distribution of quarks and gluons within the pion, and is particularly sensitive to the dynamics of the underlying strong interactions.  Investigating the properties of a hot medium in which strong interactions take place, including its dynamics, is a long-standing challenge in this field. With the application of holographic QCD methods, different form factors, such as axial-vector, electromagnetic, and gravitational form factors of hadrons has been  studied, as in Refs.~\cite{Gutsche2, Gutsche3, Gutsche4, NasibovaB, Allahverdiyev, Gutsche30, Abidin}. Furthermore, coupling constants of hadrons were reported at finite temperature in Refs.~\cite{Nasibova1, NasibovaL, NasibovaF, Taghiyeva}. The GPD of the pion has been investigated within various theoretical frameworks, including the chiral quark model \cite{Bron}, the Nambu–Jona–Lasinio (NJL) model \cite{Dorok}, double distribution approaches \cite{Davi},  and lattice QCD computations \cite{Dalley}.  

Hence, understanding how the pion form factor \cite{Nikhil, J.van},  its associated GPD \cite{Radyushkin2010, Diehl2003,  Nav}; Parton Distribution Function (PDF) \cite{Brodsky2012}; the pion charge radius \cite{ Radyushkin2010, Nikhil}; the pion mass \cite {Xuanmin}; and also pion-baryon coupling constants \cite {John} evolve with temperature is crucial for exploring the thermodynamics of hadrons in both, the low-energy regime, as well as their transition to the quark-gluon plasma beyond the confinement temperature.  The interaction between pions and other hadrons is essential not only in heavy-ion collisions in nuclear physics, but also in astrophysical scenarios, such as in the extreme-density medium of neutron stars. Temperature-dependent corrections to the pion-baryon interaction are essential for realistic modeling of proto-neutron stars, and more generally, for explaining the origins of cosmic structures, the development of galaxies, stars, and the fundamental forces of nature. 


The rest of this paper is organized as follows: \autoref{b} discusses the soft-wall model at finite temperature. \autoref{c} discusses the pion form factor, mass, pion charge radius, GPD, and the charge density of the pion at finite temperatures. In \autoref{d} we derive a temperature-dependent pion-nucleon and pion-$\Delta$ coupling constant from the bulk interaction Lagrangian between scalar and fermion fields in the AdS background. \autoref{e} presents our numerical results, and \autoref{f} concludes with a discussion.

\section{AdS/QCD Soft-Wall Model at Finite Temperature}
\label{b}

The action for the thermal soft-wall model of AdS/QCD is written as follows:
	\begin{equation}
		S=\int d^{4}x \, dr \, \sqrt{-g} \, e^{-\varphi(r,T)}L(x,r,T)  \label{1}
	\end{equation}
	where  \( g = \det(g_{MN}) \), where $g_{MN}$ is the 5D AdS metric (with $ M, \; N=0,1,2,3,5)$).  The extra dimension varies in the range $0 \leq r <   \infty $.  
	$ x=(t,\vec{x})$ is the set of Minkowski coordinates, $r$  is the Regge-Wheeler tortoise coordinate.
 The holographic thermal soft-wall model includes a dilaton field that $\varphi  (r, T)$, which depends on the temperature and the tortoise coordinate, compared to the standard soft-wall model with the  dilaton field $\varphi(z) = k^2z^2$ (where $k$ is a scale parameter of a few hundred $MeV$ and $z$ is the holographic coordinate) \cite{Gutsche2, Gutsche3} which is introduced to make the integral over $z$ finite at the IR boundary $(z\rightarrow \infty )$. In the thermal soft-wall model, the Regge-Wheeler tortoise coordinate $r$ is more convenient, since the behavior of $r$ near infinity slows down, making  calculations more controlled and allowing better handling of the infinite boundary. The two coordinates are related via:
	\begin{equation}
		r=\int \frac{dz}{f(z)}  \label{7}
	\end{equation}
	Here the thermal factor $f(z) = 1 - {z^4} / {z_H^4}$, where  $z_H$  is the position of the event horizon, which is related to the black hole Hawking temperature  $T = 1 / (\pi z_H)$. Eq. (\ref{7})   gives following relation between the  $r$ and  $z$ coordinates \cite {Gutsche2}:
	\begin{equation}
		r\approx z\left[1+\frac{z^{4}}{5z_{H}^{4}}+\frac{z^{8}}{9z_{H}^{8}}\right]  \label{8}
	\end{equation}
	The metric for the AdS-Schwarzschild spacetime in these coordinates can be written as follows: 
	\begin{equation}
		ds^{2}=e^{2A(r)}f^{\frac{3}{5}}(r)\left[dt^{2}-\frac{\left(d\vec{x}\right)^{2}}{f(r)}-dr^{2}\right] \label{9}
	\end{equation}
	where $A(r)=log(R/r)$ and $R$ is the AdS radius. Although the AdS-Schwarzchild geometry is generally used for high
$T$ (where one obtains stable solutions), for low temperatures the same metric can also be used to obtain a 
small $T$ thermal geometry via an expansion involving powers of temperature. The limit $T = 0$ corresponds to a mapping
of the AdS-Schwarzchild geometry onto the AdS-Poincare metric. At small $T$, the behavior of hadrons can be realized 
in the formalism based on the AdS-Poincare metric, using a thermal dilaton field. Thus, the following correspondence leads to equivalent results: an AdS-Schwarzchild geometry in the small $T$ limit is equivalent to an AdS-Poincare metric with a thermal dilaton field.
	
The thermal version of the usual quadratic dilaton was applied in Ref. \cite {Vega}
	\begin{equation}
		\varphi(r,T)=K^{2}(T)r^{2}
		\label{11}
	\end{equation}
	and further investigated  in Refs.~\cite{Gutsche2,Gutsche3,Gutsche4,Nasibova1}. As expected, setting $T = 0$, the thermal dilaton can be reduced to the usual dilaton  $\varphi(r,0)=k^{2}z^{2}$.
	The $K^{2}(T)$ parameter in eq. (\ref{11}) is the parameter of spontaneous breaking of chiral symmetry whose explicit form was established in Refs. \cite{Gutsche2,Gutsche3,Gutsche4,Nasibova1}:
	\begin{equation}
		K^{2}(T)=k^{2}[1+\rho(T)+O(T^{6})] 
		\label{12}
	\end{equation}
Below we shall discuss the $T$ dependence of $K$ in detail. The $\rho(T)$ function in Eq.(\ref{12}) encodes the $\emph T$ dependence of the dilaton field and was found to be of the form:
	\begin{equation} 
		\rho(T)=\frac{9\alpha\pi^{2}}{16}\frac{T^{2}}{12F^{2}}-\frac{ N_{f}^{2}-1}{N_{f}}\frac{T^{2}}{12F^{2}}-\frac{N_{f}^{2}-1}{2N_{f}^{2}}\left(\frac{T^2}{12F^2}\right)^{2} +O\left(T^6\right)  
		\label{13}
	\end{equation}
Here  $N_{f}$ is the number of quark flavors. The parameter $\alpha$ encodes the contribution of gravity to the restoration of chiral symmetry at the critical temperature $T_c$. The pion decay constant  $F$ is  proportional to the parameter $k$ as follows: $F=\frac{k\sqrt{3}}{8}$ \cite{Gasser, Gutsche2, Gutsche3}.  
The relations (\ref{12})-(\ref{13})  were obtained by introducing the thermal pre-factor $ e^{-\lambda_{T}}$ with  
 \begin{equation}
\lambda_{T}(z)=\alpha\frac{z^{2}}{z^{2}_{H}}+\gamma\frac{z^{4}}{z^{4}_{H}}+\xi\frac{k^{2}z^{6}}{z^{4}_{H}}
 \end{equation}
  in Ref. \cite{Gutsche2}. 
  The expression $\alpha \cdot  ({z^2} / {z_H^2})$ is treated as a   perturbation to the quadratic dilaton background $\varphi(z)$.
 In the holographic model, the coefficients $\alpha$, $\gamma$, and $\xi$ parametrize the thermal corrections associated with the $z^2$, $z^4$, and $z^6$ terms, respectively. Following Ref.~\cite{Gutsche2, Gutsche3}, the parameters $\gamma$ and $\xi$ are fixed to preserve gauge invariance and to keep the ground-state pseudoscalar mesons $(\pi, K, \eta)$ massless in the chiral limit,  
\[
\gamma = \frac{J(J-3)+3}{5}, \qquad \xi = 2
\]
with total spin $J$. This choice also suppresses the sixth-power radial dependence in the potential.
Thus, the thermal dilaton depends only on the parameter $\alpha$ (and not on $\gamma$ and $\xi$).


The expectation value of the scalar field denoted by $v(r, T)$, and the temperature dependent chiral condensate $\Sigma(T)$ will be  relevant to the couplings derived later in \autoref{d}.   For the finite-temperature case, the quark condensate in the expression for $v(z, T)$  depends on temperature, whereas the $z$ coordinate should be replaced by $r$. Then, the solution for $v(r,T)$ has the form \cite{Gutsche2, Gutsche3, NasibovaL, NasibovaF}: 
	\begin{equation}
		v(r,T)=\frac{1}{2}m_{q}ar+\frac{1}{2a}\Sigma(T) r^{3}   
		\label{32}
	\end{equation}
Here, \( a \) is the normalization constant. In the hard wall model, \( a = 1 \) \cite{Ahn}, but it was investigated in Ref. \cite{Cherman} in the soft wall model, and it was found that \( a = \sqrt{N_{c}}/(2\pi) \), which matches the expected \( N_c \) scaling. This result is crucial for ensuring proper behavior of QCD in the large \( N_c \) limit.
 The function $\Sigma(T)$ was determined in Ref. \cite{Gasser} by using two-loop chiral perturbation theory at finite temperature, and was subsequently applied in  Refs.~\cite{Gutsche2,Gutsche3} for the finite-temperature soft-wall model. This has the following form:
\begin{equation}
		\Sigma(T)=\Sigma \times[1-\frac{N_{f}^{2}-1}{N_{f}}\frac{T^{2}}{12F^{2}}-\frac{N_{f}^{2}-1}{2N_{f}^{2}}(\frac{T^{2}}{12F^{2}})^{2}+O(T^{6})]=\Sigma \times [1+\Delta_{T}+O(T^{6})]   
		\label{33}
\end{equation}
where $\Sigma(T=0)=\Sigma$. It can be seen that
\begin{equation}
    \Delta_{T}=-\frac{N_{f}^{2}-1}{N_{f}}\frac{T^{2}}{12F^{2}}-\frac{N_{f}^{2}-1}{2N_{f}^{2}}(\frac{T^{2}}{12F^{2}})^{2}
    \end{equation}
In the soft-wall model the dilaton field  $\varphi(r)$ is responsible for the dynamical breaking of chiral symmetry. The chiral quark condensate $\Sigma(T)$ is the result of spontaneous chiral symmetry breaking.  It was proposed in Refs.~\cite{Gutsche2,Gutsche3} that the $T$-dependence of the dilaton parameter  $K^2(T)$  should be similar to the $T$-dependence of the thermal chiral condensate:
	\begin{equation}
		K^2 (T)=k^2\frac{\Sigma(T)}{\Sigma} 
		\label{34}
	\end{equation} 
The expression for $\Sigma$ (valid at $T=0$ $GeV$) is given by
\begin{equation}
\Sigma=-N_{f}BF^{2}
\label{35}
\end{equation}
where $B$ is the condensate parameter. It was  conjectured that a similar form of the expression extends to the $T\neq0$ case. That is, at finite temperature, eq. (\ref{35}) is written as follows \cite{Gutsche2, Gutsche3}:
\begin{equation}
		\Sigma(T)= - N_{f}B(T)F^{2}(T)  
		\label{36}
\end{equation}
The temperature dependence of $F(T)$ and $B(T)$ have been studied in Ref. \cite{Gutsche2}. 
Then, following eqs. (\ref{11}) and (\ref{34}) the $T$-dependence of  $\Sigma(T)$ can be expressed in terms of the $\Delta(T)$ function \cite{Gherghetta}: 
	\begin{equation}
	\Sigma(T) = \Sigma\left[1+\Delta(T)\right]+O(T^{6})  
		\label{37}
	\end{equation} 
The temperature dependence of these quantities  ($\Sigma(T)$, $K (T)$ and $v(r, T)$) will be useful for computing several QCD observables in this work.

\section{Pion Form Factor and GPD at Finite Temperature}
\label{c}
The pion EM form factor is defined in terms of the matrix element of the pion-photon vertex, involving the pion EM current $J_\pi^\mu$. This interaction is governed by the photon-pion vertex as follows \cite{Martn}:
\begin{equation}
\mathcal{M}=\frac{1}{Q^2}\,i\,e\,\bar{u}(\kappa_2)\,\gamma_\mu\,u(\kappa_1)\,\langle \pi^{\pm}(p_2)\left|J_\pi^\mu(0)\right|\pi^{\pm}(p_1)\rangle      
\end{equation}
 Here, $e$ is the lepton electric charge, the four-vectors $\kappa_i$ and $p_i$ are momenta of leptons and pions,  respectively, and $Q=p_2-p_1$ is the virtual photon momentum.   The matrix element $\langle \pi^{\pm}(p_2)\left|J_\pi^\mu(0)\right|\pi^{\pm}(p_1)\rangle$ describes the pion-photon vertex in vacuum. Its general Lorentz structure is as follows:
\begin{equation}
 \langle \pi^{\pm}(p_2)\left|J_\pi^\mu(0)\right|\pi^{\pm}(p_1)\rangle=(p_1+p_2)^\mu\,F_\pi(Q^2)
 \label{pion-FF-M}
 \end{equation}

To compute the temperature-dependent form factor of the pion, we apply the holographic principle, which is the correspondence between the generating function of QCD and the generating function of AdS gravity $(Z_{AdS}(T)=e^{iS_{int}(T)}=Z_{QCD}(T))$ at finite temperature. Here, the thermal action $S_{int}(T)$ expressing the interaction between the bulk fields takes the following form: 
\begin{equation}\label{eff-action}
S_{int} = \int d^4x dr\,\sqrt{-g} g^{mn} \, \phi_m(x,r,T) \left[ X_{p_1}(x,r,T) \, \partial_m X^*_{p_2}(x,r,T) - X^*_{p_2}(x,r,T) \, \partial_m X_{p_1}(x,r,T) \right]
\end{equation} 
where,  $X(x,r, T)$  and  $\phi_m(x,r, T)$  are the fields dual to the pion and the virtual photon at finite temperature. The bulk field $X(x,r,T)$ dual to the pion can be expressed as thermal waves in AdS spacetime (following the temperature-independent case in \cite{Martn}):
\begin{eqnarray*}
X(x,r,T)&=&e^{-i\,p\cdot x}\,P(r, T) 
\end{eqnarray*}
 where $P(r, T)$ is the temperature-dependent profile function of the pion. To obtain the photon propagator, we express the four components of the field dual to the virtual photon $\phi_\mu (x,r, T)$ as a Fourier transform: 
\begin{eqnarray*}
\phi_\mu (x,r, T)&=& \int dQ \,  \eta_\mu\,e^{-i\,Q\cdot x}\,   V(Q, r, T)     
\end{eqnarray*}
where $V(Q, r, T)$ is the   
vector non-normalizable mode dual to the virtual photon. This is the photon bulk-to-boundary propagator, whose solution shall be determined in what follows.  Here $\eta_\mu$ is the polarization 4-vector of the photon.   

The pion form factor at finite temperature is obtained analogously to the zero-temperature case discussed in \cite{Martn}. This yields the following result:  
\begin{equation} 
  F_{\pi}(Q^2, T) = \int dr\,e^{3\,A(r)}P(r, T)\, V(Q, r, T) \,P(r, T)  
\end{equation}
Substituting Schrodinger-like modes for the bulk scalar field $P(r, T)=e^{-3\,A(r)/2}\,f_{0}(r,T)$, the above expression for the pion form factor can be written as:
\begin{equation}
F_\pi(Q^2,  T)=\int{dr\,f^{*}_{0}(r, T)\, V(Q, r, T)   \,f_{0}(r, T)}     \end{equation}
 The expression for the photon bulk-to-boundary propagator $V(Q, r, T)$ can be obtained using the EOM for the vector field with specific boundary conditions. For the finite temperature case, in the soft-wall model with a thermal dilaton, we use the formal replacements $z \rightarrow r, \; k \rightarrow K(T) $, which do not change the EOMs. Consequently, the known expression for the vector  propagator at finite temperature is obtained in Ref. \cite{Gutsche2}. In the soft-wall model with a thermal dilaton field $\varphi (r, T)$ defined in eq. (\ref{11}), the EOM for the vector field, corresponding to the coordinate $r$, is expressed as  \cite{Gutsche2, Gutsche3}:
	\begin{equation}
		\partial_{r}\left(\frac{e^{-\varphi(r, T)}}{r}\partial_{r}V(Q, r, T)\right)- Q^{2}\frac{e^{-\varphi(r, T)}}{r}\partial_{r}V(Q, r, T)=0 
		\label{14}
	\end{equation}
The propagator $V(Q,r,T)$ has to satisfy  the boundary condition $V(Q, r=0, T)=1$.
	 Eq. (\ref{14}) is similar to the one at zero temperature, and the only difference between them is the $T$-dependence of the dilaton parameter. 
	Hence, we have the solution of eq. (\ref{14}) written as in \cite{Gutsche3}:  
\begin{eqnarray}
V(Q, r, T) &=& \Gamma \left(1+a(Q, T)\right)U\left(a(Q, T), 0, K^{2}(T)r^{2}\right)  \nonumber \\ 
&=& K^{2}(T) \; r^{2}  \int_{0}^{1}\frac{dx}{(1-x)^{2}}x^{a(Q, T)}e^{-K^{2}(T)r^{2}\frac{x}{1-x}}   
		\label{15}
\end{eqnarray}
 Here $\Gamma \left(1+a(Q, T)\right)$ is Euler's gamma function and $a(Q,T)=\frac{Q^{2}}{4K^{2}(T)}$. $U\left(a(Q, T), 0, K^{2}(T)r^{2}\right)$ is a Tricomi function, also known as a confluent hypergeometric function of the second kind.

Now, $f_0(r, T)$, which appears in the finite temperature pion profile function  can be written in the following form, corresponding to the zero-temperature case by replacing $k \to K(T)$ and $z \to r$   \cite{Martn}:
\begin{equation}
    f_0(r, T)=\sqrt{\frac{2\,K^4(T)}{R^3}}\,r^3 
\end{equation}


With all the above ingredients, the expression for the pion form factor at finite temperature takes an analytic form analogous to that of the zero-temperature case \cite{Martn}:
\begin{eqnarray}
F_{\pi}(Q^2, T)&=&R^3\,\int_0^\infty{dr\,\frac{e^{-K^2(T)\,r^2}}{r^3}\,f_0^*(r, T)\, V(Q,r, T)   \,f_0(r. T)}\\
&=&2\,K^6(T)\,\int_0^1{\frac{dx}{(1-x)^2}\,x^{\frac{Q^2}{4\,K^2(T)}}\,\int_0^\infty{dr\,r^5\,e^{-\frac{K^2(T)\,r^2}{1-x}}}}\\
&=&\frac{32\,K^4(T)}{\left(Q^2+4\,K^2(T)\right)\left(Q^2+8\,K^2(T)\right)} \label{FFF}
\end{eqnarray}
Here, $x$ is the parton momentum fraction, $K(T)$ represents the dilaton slope that defines the scalar meson linear Regge trajectory at finite temperature, and $R=1$ is the AdS radius of curvature.   


Furthermore, the mass of mesons in the chiral limit at finite temperature was determined in \cite{Gutsche2, Gutsche3} as follows: 
\begin{equation}  
M^2_{(n)}(T) = 4 K^2 (T)  (n + \frac{m+J}{2}) + \frac{4 \pi^4 T^4}{5 k^2} J (J + 1) [ (m + 1)(m+2) + (6 n - J)(n+m+1) - nJ  ]  
\label{masschl}
\end{equation}
Here  $J$ is the spin of the meson, $n$ is the radial quantum number, and $m = N+L-2$. For the pion, $N=2$ is the  parton number, $n = 0$ and $l=0$. The latter corresponds to the orbital angular momentum $L=l(l+1)$. Since eq. (\ref{masschl}) was derived in the chiral limit, where quark masses vanish, one obtains $M^2_{(0)}(T) = 0$ for the pion. 
Leading order corrections to the above mass formula come from considering the effects of chiral symmetry breaking \cite{GutscheChs}. We will compute this contribution to the pion mass using a temperature-dependent extrapolation of the Gell-Mann-Oakes-Renner (GMOR) relation \cite{GMOR} (see also \cite{Gutsche2}): 
\begin{eqnarray}
    M_{\pi}^2 (T) = - \frac{(m_u + m_d) \, \Sigma (T) }{F^2 (T)}
\label{gmor}
\end{eqnarray}
where the temperature dependence of the quark chiral condensate $\Sigma (T)$ in the thermal soft-wall model follows from eq. (\ref{37}) \cite{Gutsche2}. The thermal profile of the chiral condensate has independently been estimated using the hadron resonance gas model  \cite{janko}, and in chiral perturbation theory \cite{gomezn}). The temperature dependence of the meson decay constant $F (T)$ was derived at one-loop in \cite{Gasser} and takes the form 
\[ F (T) = F \left( 1 - \frac{N_f}{2} \frac{T^2}{12 \, F^2}  + {\cal O} (T^4) \right)  \]
Both $\Sigma (T)$ and $F (T)$ depend on the flavor number $N_f$. The GMOR relation itself holds only up to leading order in chiral perturbation theory, and is subject to higher-order corrections. Nevertheless, as our results below show, eq. (\ref{gmor}) gives a good estimate of the pion mass at zero temperature.

Next, using the expression for the EM form factor of the pion (eq.(\ref{FFF})), we compute the mean-square charge radius of the pion at finite temperature as follows:  
\begin{equation}
    \langle r^2_{\pi}(T) \rangle  = - 6 \,\left.\frac{d F_{\pi}(Q^2, T)}{d Q^2}\right|_{Q^2=0}   
\label{raio}
\end{equation}

The GPD of the pion at zero temperature has been studied earlier in  \cite{Hashamipour, Hashamipour1, Hashamipour2}. Here, we will consider the finite temperature case at zero skewness.
As with the radius, the GPD of the pion at finite temperature can also be determined from the thermal pion form factor. The thermal GPD depends on four varibales:  the momentum transfer $Q^2$, the Bjorken scaling variable $x$, the skewness $\zeta$, and temperature $T$. Here, $Q^{2}$ is the square of the four momentum transfer between the hadrons, the variable $x$ stands for the light-front longitudinal momentum fraction, and \( \zeta \)  quantifies the longitudinal momentum  between initial and final states.

The general expression for the spin non-flip GPD is defined through the off-forward matrix elements of the bilinear vector current as in  \cite{Nav}:
\begin{equation}
H(x, \zeta, Q^2) = \int \frac{dy^-}{8\pi} e^{i x P^+ \frac{y^-}{2}} \langle P' | \bar{\psi}(0) \gamma^+ \psi(y) | P \rangle \bigg|_{y^+ = 0, \mathbf{y}_\perp = 0}
\end{equation}
Where $Q = P'-P$ with \( P \) and \( P' \), which indicate the initial and final momenta of a pion with mass \( M \), respectively. The quark field operators \( \psi(0) \) and \( \psi(y) \) are evaluated at two separate points in spacetime, 0 and \( y \).   
When $\zeta \neq 0$, the GPD quantifies the spatial and momentum structure of hadrons. In processes like Deeply Virtual Compton Scattering (DVCS), $\zeta$ is small (typically $0.1$ - $0.4$). 


In the special case when the momentum transfer $Q^2$ and the skewness $\zeta$ are zero, the GPD reduces to the parton distribution function (PDF).  The PDF only describes the probability density of finding a parton (quark or gluon) carrying a certain fraction \( x \) of the hadron's total momentum, and is denoted by $f(x)$ as follows:  \begin{equation}
    H(x, \zeta = 0, Q^{2} = 0) = f(x)
\end{equation}

The GPD of the pion \( H^\pi(x, \zeta, Q^2) \) considered here refers to the unpolarized GPD (distribution of partons without considering spin correlations) \cite{Azizi}. 
The relation between the pion form factor (eq. \ref{FFF}) and pion GPD is as follows:
\begin{equation}
F(Q^2, T) \;   =  \int^{1}_{0} dx \,  H(x, \zeta, Q^2, T)
\end{equation}
 With this, we obtain the following temperature-dependent expression for the pion GPD
\begin{equation}
    H(x, \zeta=0, Q^2, T) = 2 K(T)^6 \cdot \frac{x^{\frac{Q^2}{4K(T)^2}}}{(1 - x)^2} \cdot \int_0^{\infty} r^5 \exp\left( -\frac{K(T)^2 r^2}{1 - x} \right) \, dr
\end{equation}

Pions also possess an charge density. This density exists only for charged pions, and not for neutral pions \cite{Nav}. The EM density of the pion at finite temperature is obtained from the pion EM form factor as follows: 
\begin{equation}
    \rho_{\text{EM}}^{\pi^{\pm}}(b,T) \, = \, \pm \frac{1}{2\pi} \int_0^\infty dQ  \,  J_0 (Qb) \, F_\pi(Q^2,T)
\end{equation}
where $ J_0 (Qb) $ is the Bessel function of the first kind (order 0).  The charge density here is taken at the transverse distance  (impact parameter) $b$ and finite temperature $T$.  The parameter \( b \) is defined as the magnitude of the two-dimensional vector \(\mathbf{b}\) in the transverse plane,
\[
b = |\mathbf{b}| = \sqrt{b_x^2 + b_y^2},
\]
where \( b_x \) and \( b_y \) are the components of the transverse vector \(\mathbf{b}\) along the \( x \)- and \( y \)-axes, respectively.

\section{Pion-Baryon Coupling Constants at Finite Temperature}
\label{d}

\begin{figure}[htbp]
    \centering
    \includegraphics[scale=0.79]{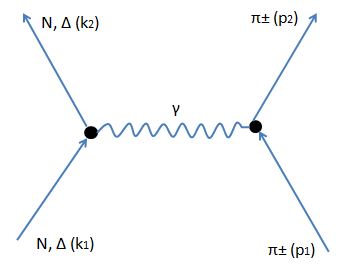}
   
     \caption{  The nucleon or $\Delta$ baryon interaction with the pion via the exchange of a virtual photon. } 
    \label{2a}
\end{figure}

We now derive expressions for the pion-nucleon-nucleon, and  pion-$\Delta$-$\Delta$  coupling constants at finite temperatures within the thermal soft-wall model of holographic QCD.  Fig. \ref{2a}  represents the Feynman diagram of a scattering between a pion and nucleon ($\Delta$ baryon)
 through the exchange of a virtual photon.
\FloatBarrier

We begin with the bulk pion-nucleon interaction Lagrangian corresponding to the zero temperature case  \cite{Ahn}. This is expressed via the  
minimal gauge coupling and the Yukawa coupling with scalars as follows:   
\begin{eqnarray}
     {L}^{(0)}_{\pi NN}(x,r,T) = \bar{N}_1(x,r,T) \Gamma^5 P(x,r,T) N_1(x,r,T) - \bar{N}_2(x,r,T) \Gamma^5 P(x,r,T) N_2(x,r,T)  \nonumber \\
     -g_{Y} \left( \bar{N}_1(x,r,T) X(x,r,T) N_2 (x,r,T)+ \text{h.c.} \right)  \hspace{3.4cm}
\label{p30}
\end{eqnarray}


Here $\Gamma^5 = -i \gamma^5$ is the fifth Dirac matrix corresponding to $AdS$ spacetime.  
The interaction terms here involve gauge fields, bulk scalars and nucleons. $g_Y$ denotes the Yukawa coupling. Notice that this Lagrangian takes the same form as the zero temperature case, but now with a $T$-dependence in the fields  \cite{Ahn}. 

Next, a Kaluza-Klein (KK) expansion is performed for the transverse four-dimensional components of the AdS bulk pion field $P(x, r, T)$.  In the unitary gauge, the thermal pion is identified with its lowest KK mode  
\begin{equation}
    P(x,r,T) = c_\pi \, P_{0} (x) \, f_0(r,T) ,
    \label{eq:pion_field}
\end{equation}
where $c_\pi$ is a normalization constant, and $P_0 (x)$ is  the Kaluza-Klein (KK) $n = 0$ mode dual to the pseudoscalar pion. The function $f_0(r, T)$ is the extra-dimensional temperature-dependent wave function (or profile function) of the pion. This mode of the KK expansion (that is, the ground state $n=0$) governs the low-energy effective dynamics. 
We also include the following (parity-invariant) temperature-dependent magnetic gauge coupling in the bulk to the minimal gauge interaction as follows:
\begin{eqnarray}
    L^{(1)}_{\pi NN} = &i\kappa_1& \left( \bar{N}_1 \Gamma^{MN} (F_L)_{MN} N_1 - \bar{N}_2 \Gamma^{MN} (F_R)_{MN} N_2 \right) 
    + \frac{i}{2} \kappa_2 \left( \bar{N}_1 X \Gamma^{MN} (F_R)_{MN} N_2  \right. \nonumber \\
  &\,&  \left.  +  \bar{N}_2 X^{\dagger} \Gamma^{MN} (F_L)_{MN} N_1 \right) - \text{h.c.}  
\label{p42}
\end{eqnarray}

 Here the  \(\Gamma^{MN}\) is antisymmetric combinations of Dirac gamma matrices, defined as
$
\Gamma^{MN} = \frac{1}{2}[\Gamma^{M}, \Gamma^{N}]$ in AdS space. In a 5-dimensional holographic model the indices
\( M, N = 0, 1, 2, 3, 5 \)
denote the five spacetime dimensions.

These operators act as generators of Lorentz transformations in the spinor representation and commonly appear in fermionic couplings to gauge field strength tensors with corresponding Lorentz indices.

The temperature-dependent 5-dimensional spinors \( N_1(x, r, T) \) and \( N_2(x,r, T) \)  are expressed in terms of 4-dimensional field operators $u_L (p)$, and $u_R (p)$, and  are written via their respective finite-temperature Fourier transforms as follows (Ref. \cite{Ahn}):
\begin{eqnarray}
    N_1 (x,r,T) =N_{1L} (x,r,T) + N_{1R} (x,r,T)= \frac{1}{(2\pi)^4} \int d^4 p' \, e^{-ipx} \left[ F_{1L} (r,T) u_L (p) + F_{1R} (r,T) u_R (p) \right]  
   \nonumber\\
N_2 (x,r,T) = N_{2L} (x,r,T) + N_{2R} (x,r,T)=\frac{1}{(2\pi)^4} \int d^4 p' \, e^{-ipx} \left[ F_{2L} (r,T) u_L (p) + F_{2R} (r,T) u_R (p) \right]  \qquad
\label{p32}
\end{eqnarray}
Here, \( F_{1L/2L}(r,T) \) and \( F_{1R/2R}(r,T) \) are the left and right profile functions for nucleons at finite temperature.

Similarly, the spinors \( \bar{N}_1(x,r,T) \) and \( \bar{N}_2(x,r,T) \) are expressed as:
\begin{eqnarray}
\bar{N}_{1} (x,r,T) = \bar{N}_{1L} (x,r,T) + \bar{N}_{1R} (x,r,T) =\frac{1}{(2\pi)^4} \int d^4 p' \, e^{ip'x} \left[ F_{1L}^* (r,T) \bar{u}_{L} (p') + F_{1R}^* (r,T) \bar{u}_{R} (p') \right]  \nonumber\\
\bar{N}_{2} (x,r,T) = \bar{N}_{2L} (x,r,T) + \bar{N}_{2R} (x,r,T) =
\frac{1}{(2\pi)^4} \int d^4 p' \, e^{ip'x} \left[ F_{2L}^* (r,T) \bar{u}_L (p') + F_{2R}^* (r,T) \bar{u}_R (p') \right]  \qquad
\label{p33}
\end{eqnarray}  



To derive the temperature-dependent pion-nucleon coupling constant  we consider the following bulk interaction action in  $AdS$, which involves the thermal nucleon interacting
with the thermal pion:
\begin{equation}
S_{int}(T) = \int d^4 x \, dr \, e^{-\phi(r,T)} \sqrt{g} \, L^{AdS}_{\pi NN}(x,r,T) 
\label{e35}
\end{equation}
where 
\[ L^{(AdS)}_{\pi NN}(x,r,T) = L_{\pi NN}^{(0)} (x,r,T)+ L_{\pi NN}^{(1)}(x,r,T)   \]


We then invoke the holographic correspondence \cite{Maldacena}.  Adapted to the the context of AdS/QCD, this posits that the generating functional $Z_{\text{AdS}}$ of the gravitational theory in the bulk is identical to the generating functional $Z_{\text{QCD}}$ of QCD on the UV boundary of this space-time
\begin{equation}
    Z_{\text{AdS}}=e^{iS_{int}}=Z_{\text{QCD}}
\label{p36}
\end{equation}

To find the matrix element of the current operator, which interacts with the pion in the boundary QCD theory, we take the variation of the bulk generating functional \( Z_{\text{AdS}} \) with respect to the bulk pion field \( \mathcal{P}_a(q) \), evaluated at the boundary value of the pion 
\begin{equation}
\langle J_a(q, T) \rangle = -i \, \left. \frac{\delta Z_{\text{AdS}}(T)}{\delta \mathcal{P}_a (q)} \right|_{\mathcal{P}_{a} = 0}
\label{p38}
\end{equation}
Here the index "$a$" denotes isospin degrees of freedom. 

The variational derivative above is calculated using explicit expressions of the thermal profile functions of the bulk fields. We substitute the relations for the Fourier transformation of the nucleon field (eqs. (\ref{p32}) and (\ref{p33})), and the creation and annihilation operators into the Lagrangian $L^{AdS}_{\pi NN}(x, r, T)$. Using this explicit form of $L^{AdS}_{\pi NN}(x,r,T)$ in the action eq. (\ref{e35}), yields the analytic expression for $Z_{\text{AdS}}(T)$.  Terms of the thermal action are computed in momentum space. Then, for each Lagrangian term, we get the following contribution to the nucleon current, expressed in terms of spinors $u(p)$ and $\bar{u}(p')$:  
\begin{equation}
\langle J_a(q, T) \rangle  = g_{\pi NN}(T) \, \bar{u}(p') \, \gamma^5 \, \tau_a \, u(p) 
\label{p39}
\end{equation}
where, $\tau_a$ are the Pauli matrices,  \( p' \) is the 4-momentum of the nucleon field before the interaction at finite temperature, and \( p \) is the momentum after the interaction. The energy-momentum conservation law between the 4-dimensional momenta \( p' \), and \( p \) is given by \( q = p' - p \).  In the above, $g_{\pi NN} (T)$ denotes the thermal pion-nucleon coupling constant.


%
%
%

The finite-temperature pion-nucleon coupling \( g_{\pi NN}^{(0)}(T) \) associated to the 
 $L^{(0)}_{\pi NN}(x,r,T)$ part of the Lagrangian is then as follows: 
\begin{eqnarray}
    g^{(0)}_{\pi NN}(T) = \int_0^{\infty} \frac{{\rm d}r}{2\sqrt{2}r^4} 
\left[ f_0(r, T) \left( F^{(n)*}_{1L} (r,T) F^{(m)}_{1R}(r,T) - F^{(n)}_{2L}{}^*(r,T) F^{(m)}_{2R}(r,T) \right) \right.  \nonumber \\
\left. \times \frac{g_{Y}r^2}{2vg_5^2} \, \partial_r \left( f_0(r,T) \right) \left( F^{(n)}_{1L}{}^*(r,T) F^{(m)}_{2R}(r,T) - F^{(n)}_{2L}{}^*(r,T) F^{(m)}_{1R}(r,T) \right) \right]
\label{p41}
\end{eqnarray}
Here $g_5^2 = \frac{12 \pi^2}{N_c}$ and $N_c$ is the color number. This is the general expression for the pion-nucleon coupling constant (corresponding to $L^0_{AdS}(x,r,T))$, which includes ground states and excited states of nucleons. In our numerical results, we will investigate the pion-nucleon coupling constant in the ground state of the nucleons, where $n=m=0$. Analogously, one can determine the pion-nucleon-Roper coupling, associated to the excited states of the nucleon ($n \neq m \neq 0$) by using the appropriate form of the nucleon profile function $F^{(n)*}_{L/R} (r,T)$ and the corresponding nucleon mass of this state. 
For the nucleons, the total angular momentum $L$ is $0$ and spin $\emph{J}$ is equal to $\frac{1}{2}$ \cite{Azizi2}. The explicit form of the 
profile function for nucleons at finite temperature is as follows \cite{Gutsche2, Gutsche3, Nasibova1, NasibovaL, NasibovaF}:
\begin{eqnarray}
F_{L/R}^{(n)}(r,T)=\sqrt{\frac{2\Gamma (n+1)}{\Gamma (n+m_{L}/m_{R}+1)}}K^{m_{L}/m_{R}+1}r^{m_{L}/m_{R}+\frac{1}{2}}e^{-\frac{K^{2}r^{2}}{2}}L_{n}^{m_{L}/m_{R}}\left(K^{2}r^{2}\right)  
\label{43}
\end{eqnarray}
This profile function obeys the normalization condition given in \cite{Gutsche2, NasibovaB} as follows:
\begin{equation}
		\int_{0}^{\infty } dr F^{(m)}_{L,R}(r,T) F^{(n)}_{L,R}(r,T)=\delta_{mn}.
		\label{43norm}
	\end{equation}
Here $L_{n}^{m_{L}/m_{R}}\left(K^{2}r^{2}\right)$ is the Laguerre polynomial and $\Gamma(n)$ is the gamma function.
 
For nucleons, the parton number $N=3$ and the orbital angular momentum  $L=0$. In addition,  
$m_{L/R}=m\pm\frac{1}{2}$; therefore, for left and right nucleons $m_{L}=2$ and $m_{R}=1$ respectively.
For parity-even states, the eigenfunctions have the properties: 
\[ F^{(n)}_{1L}(r,T) \, = \, F^{(n)}_{2R}(r,T)  \qquad \mbox{and} \qquad  F^{(n)}_{1R}(r,T) \, = \, -F^{(n)}_{2L}(r,T)  \]
whereas parity-odd states satisfy: 
\[ F^{(n)}_{1L}(r,T) \, = \, -F^{(n)}_{2R}(r,T) \qquad \mbox{and} \qquad 
 F^{(n)}_{1R}(r,T) \, = \, F^{(n)}_{2L}(r,T)  \]
The lowest-lying ($n=1$)  nucleons, are parity even, and excited resonances appear as parity-doublets $(N^+,N^-)$ with the parity-even states ($N^+$) being slightly lighter than the parity-odd states ($N_-$). These we have verified numerically.



Using the Kaluza-Klein (KK) decomposition for the five-dimensional spinors as before, the additional contribution to pion-nucleon couplings can be derived as
\begin{eqnarray}
    g^{(1)}_{\pi NN} (T) = -2m^{(mn)} \int_{0}^\infty \frac{dr}{r^3} f_0 (r,T) \left[ \kappa_1 \left( F^{(n)*}_{1L}(r,T) F^{(m)}_{1L}(r,T) + F^{(n)*}_{2L}(r,T) F^{(m)}_{2L}(r,T) \right) \right. \nonumber \\
   \left. - \kappa_2 \, v(r,T) \left( F^{(n)*}_{1L}(r,T) F^{(m)}_{2L}(r,T) - F^{(n)*}_{2L}(r,T) F^{(m)}_{1L}(r,T) \right) \right]  \quad  
\label{p43}
\end{eqnarray}
 where $\kappa_{1}$ and $\kappa_{2}$ are free parameters, whose values will be determined in the next section (in  the numerical results).  The second term in eq. (\ref{p43}) will vanish due to its antisymmetric structure after substituting the above-mentioned relations between parity-even and -odd states. Hence, the parameters $k_2$  does not contribute to the analysis and can be safely omitted.  Here $m^{(mn)} = m^{(m)}_N + m^{(n)}_N$, and $m^{(m)}_{N}$ and $m^{(n)}_{N}$ denote nucleon masses in the ground and excited states, respectively.   


Putting together the contributions from eqs. (\ref{p41}) and (\ref{p43}), yields our result for the finite temperature pion-nucleon-nucleon coupling derived within the context of the thermal soft-wall model:
\begin{eqnarray}
 g^{}_{\pi NN}(T)=g^{(0)}_{\pi NN}(T) + g^{(1)}_{\pi NN}(T)    
\end{eqnarray}

Next, we investigate the temperature dependence of the pion-$\Delta$-$\Delta$  coupling. This can be derived from bulk interaction terms involving gauge fields and $\Delta$ resonances 
\begin{equation}
{\cal L}_{\pi \Delta \Delta} =
\bar\Psi_1^{\mu} \Gamma^r P_r \Psi_{1\mu} -\bar\Psi_2^{\mu} \Gamma^r
P_r \Psi_{2\mu} - g_{3/2} (\bar\Psi_1^{\mu} X^3 \Psi_{2\mu} + \text{h.c.})\,.
\end{equation}
where $\bar\Psi_{1/2}^{\mu}$ are  $\Delta$ baryon fields and $X$ is the scalar field.  Using this we obtain  part of the four-dimensional pion-$\Delta$-$\Delta$ coupling constant corresponding to bulk interaction terms in ${\cal L}_{\pi \Delta \Delta}$ as follows   
\begin{eqnarray}
g^{(0)nm}_{\pi \Delta \Delta}(T) &=& \! -\int^{\infty}_0 \frac{dr}{2r^2}
\left[f_0(r,T)\Big(f^{(n)*}_{1L}(r,T) f^{(m)}_{1R}(r,T) - f^{(n)*}_{2L}(r,T)
f^{(m)}_{2R}(r,T)\Big)\right.\nonumber\\
&&\quad \left. -\frac{3g_{Y} r^2v(r,T)}{2
g^2_5}\left(\frac{f_0(r,T)}{r}\right)^\prime \Big(f^{(n)*}_{1L}(r,T)
f^{(m)}_{2R}(r,T) - f^{(n)*}_{2L}(r,T) f^{(m)}_{1R}(r,T)\Big)\right] \qquad 
\end{eqnarray}
The left/right $\Delta$  baryon profile functions  $f^{(n)}_{L/R}$ obtained from the Rarita-Schwinger equation are as follows \cite {NasibovaL}:
\begin{eqnarray}
f_{L/R}^{(n)}(r,T)=\sqrt{\frac{2\Gamma (n+1)}{\Gamma (n+m_{L}/m_{R}+1)}}K^{m_{L}/m_R+1}r^{m_{L}/m_R+\frac{1}{2}}e^{-\frac{K^{2}r^{2}}{2}}L_{n}^{m_{L}/m_R}\left(K^{2}r^{2}\right)  
\end{eqnarray}
where $f_{L/R}^{(n)}(r,T)$ obey a normalization condition analogous to that in eq. (\ref{43norm}).

Additional contributions to the 
pion-$\Delta$-$\Delta$ couplings arise from the  magnetic coupling in the bulk, similar to the couplings with (excited) nucleons at finite temperature,
\begin{eqnarray}
{\cal L}_{F\Delta \Delta} &=& i \kappa_3 \Big[\bar\Psi_1^M \Gamma^{NP}
{(F_L)}_{NP} \Psi_{1M} - \bar\Psi_2^{M} \Gamma^{NP} {(F_R)}_{NP}
\Psi_{2M}\Big] \nonumber\\
&& \, +\frac{i}{2} \kappa_4 \Big[\bar\Psi_1^M X^3\Gamma^{NP} {(F_R)}_{NP}
\Psi_{2M} + \bar\Psi_2^{M} (X^{\dagger})^3\Gamma^{NP} {(F_L)}_{NP}
\Psi_{1M}-\text{h.c.}\Big]  
\end{eqnarray}
Here, \(\Gamma^{NP}\) is the antisymmetric combination of Dirac gamma matrices in curved space, defined as \(\Gamma^{NP} = \frac{1}{2}[\Gamma^N, \Gamma^P]\), where the indices \(N, P = 0, 1, 2, 3, 5\) in AdS space.
Using the KK mode decomposition for the spinors as before, the 
additional contribution to the pion-$\Delta$-$\Delta$ coupling constant at finite temperature is obtained as 
\begin{eqnarray}
g^{(1)nm}_{\pi \Delta \Delta}(T) =
-(m^{(n)}_{\Delta}+m^{(m)}_{\Delta})\int^{\infty}_0 \frac{dr}{r} f_0(r,T)
\left[\kappa_3\Big(f^{(n)*}_{1L}(r,T) f^{(m)}_{1L}(r,T) + f^{(n)*}_{2L}(r,T)
f^{(m)}_{2L}(r,T)\Big)\right.\nonumber\\
 \left. -\kappa_4 \,  (v(r,T))^3\Big(f^{(n)*}_{1L}(r,T) f^{(m)}_{2L}(r,T)
- f^{(n)*}_{2L}(r,T) f^{(m)}_{1L}(r,T)\Big)\right] \quad 
\label{53}
\end{eqnarray}
where $\kappa_3$ and $\kappa_4$ are free parameters, whose values will be determined in the next section (in the  numerical results). Once again, the second term in eq. (\ref{53}) will vanish due to its antisymmetric structure after substituting the  relations between parity-even and -odd states. Therefore, the parameter $k_4$  does not contribute to the coupling. 

The total coupling constant is thus given by the sum of the above two contributions as follows:
\begin{eqnarray}
 g^{nm}_{\pi \Delta \Delta}(T)=g^{(0)nm}_{\pi \Delta \Delta}(T) + g^{(1)nm}_{\pi \Delta \Delta}(T)    
\end{eqnarray}

\section{Numerical Results}
 \label{e}

The thermal soft-wall model
includes a parameter $k$, which is fixed by the pion decay constant. Following  Refs.~\cite{Gutsche2, Gutsche3}  we take  $k=0.402$ GeV, when $F=0.087~GeV$. This will be used in our calculations below.

\subsection{Pion Electromagnetic Form Factor}

\begin{figure}[htbp]
    \centering
    \includegraphics[scale=0.40]{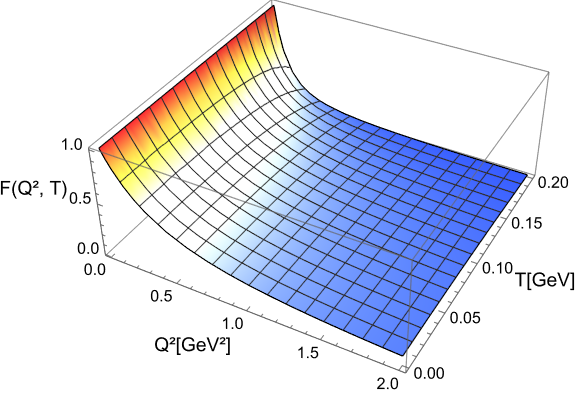} \hspace{1em}
    \includegraphics[scale=0.41]{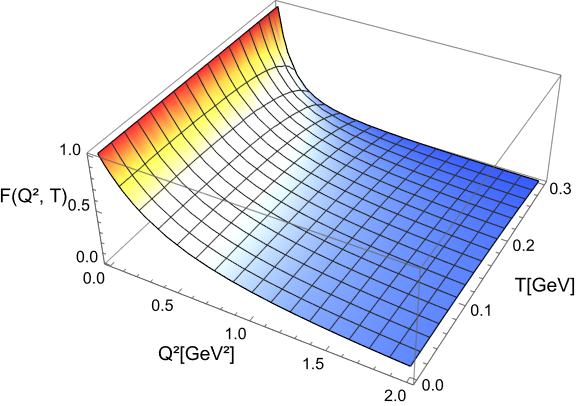}
    \par{$a)$  $\alpha=0$} \hspace{8em} {$b)$ $\alpha=0.2$}\par\vspace{1em}

    \includegraphics[scale=0.43]{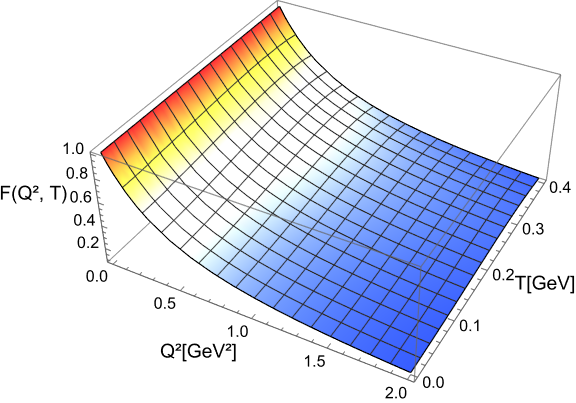}
    \par{$c)$  $\alpha=0.4$}\par\vspace{0.5em}

    \caption{$T$ and $Q^{2}$ dependence of the pion form factor $F_{\pi}(Q^2,T)$  for different value of parameter $\alpha$. }
    \label{fig:formfactor}
\end{figure}

In Fig. \ref{fig:formfactor}, we present $3D$ graphs of the temperature-dependent form factor \( F_{\pi}(Q^2, T) \) at $\alpha=0$, $\alpha=0.2$, and $\alpha=0.4$ (recall that $\alpha$ reflects the influence of gravitational dynamics). These plots show that with increasing values of $\alpha$, the pion form factor also increases.  The influence of gravitational dynamics increases the significance of chiral symmetry restoration with rising temperature. Note that these figures show that the critical temperature increases with higher values of $\alpha$.

\begin{figure}[htbp]
    \centering
   \includegraphics[scale=0.55]{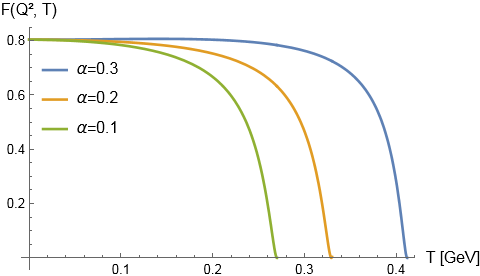}
   \hspace{1em}
   \includegraphics[scale=0.55]{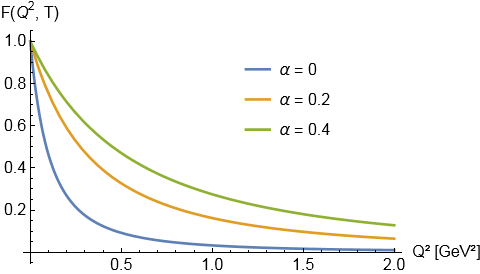}
    \par{$(a)$  } ${Q^{2}=0.1}$ $GeV^2$ \hspace{12em} {$(b)$  $T=0.2$ $GeV$}\par\vspace{1em}
      \caption{ (a): $T$ dependence of the pion form factor $F_{\pi}(Q^2, T)$   and (b): $Q^{2}$ dependence of the $F_{\pi}(Q^2, T)$  for different values of $\alpha$.   }
    \label{FF3a}
\end{figure}

     \FloatBarrier
Fig. \ref{FF3a} shows 2D illustrations of the combined dependence of \( F_{\pi}(Q^2, T) \) on both temperature and momentum squared for different values of \( \alpha \).    In Fig. \ref{FF3a} (a) We observe a significant increase in the form factor at a given temperature with increasing values of \( \alpha \). Here \( Q^2 = 0.1~GeV^2\) is fixed.  We observe that the form factor, which depends on temperature, increases as \( \alpha \) increases. This indicates that gravitational dynamics (represented by \( \alpha \)) have a strong influence on the internal structure of hadrons and their form factor. 
The plot in Fig. \ref{FF3a} (b)  shows that the profile of the form factor versus $Q^2$ when $T = 0.2~GeV$. The value of the form factor increases consistently with increasing $\alpha$.

\FloatBarrier

\begin{figure}[htbp]
    \centering
  \includegraphics[scale=0.6]{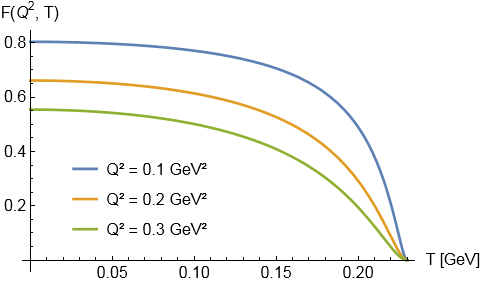}
   \hspace{1em}
   \includegraphics[scale=0.6]{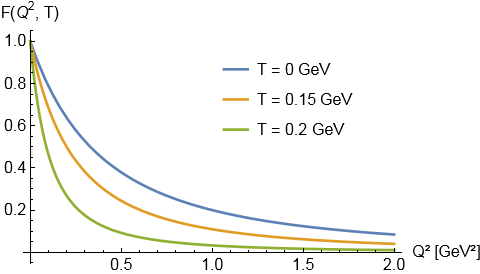}
    \par\textbf{$(a) ~ \alpha = 0$ } \hspace{10em} \textbf{$(b)  ~ \alpha = 0$ }\par\vspace{1em}
      \caption{ (a): $T$ dependence of the pion form factor $F_{\pi}(Q^2, T)$ for different values of $Q^{2}$, and (b): $Q^2$ dependence of the pion form factor at different values of $T$.  } 
    \label{3a}
\end{figure}

In Fig. \ref{3a}, we show the temperature dependence of the pion form factor \( F_{\pi}(Q^2, T) \) at various values of the momentum transfer squared \( Q^{2} = 0.1~GeV^2, ~0.2~GeV^2, ~0.3 ~GeV^2 \), and also the dependence of of the form factor on momentum squared \( Q^2 \) at different temperatures $T = 0~GeV, ~0.15~GeV, ~0.2 ~GeV$. Here $\alpha = 0$ is fixed.  From this, it is evident that the temperature dependence of the pion's form factor decreases as the momentum increases. We also observe that with an increase in temperature the pion form factor decreases. 

\FloatBarrier

Now, let us contrast the above with  experimental values. The most commonly used method for experimentally measuring the pion form factor is electron-pion scattering.  The values of the experimental pion form factor from international collaborations are shown in Table \ref{datatable}. Experiments as well as theoretical predictions indicate that the pion form factor decreases asymptotically as \( Q^2 \) increases, i.e. the form factor behaves approximately as  \cite{Martn}:
\[
F_\pi(Q^2) \sim \frac{1}{Q^2}  
\]
This behavior suggests that the interaction between quarks inside the pion weakens at higher \( Q^2 \) values. This phenomenon is understood in QCD in terms of how the strong interaction becomes weaker as the momentum transfer increases, leading to the observed asymptotic form factor behavior that we observe above.

\begin{table}[htbp]
    \centering
\begin{tabular}{|l|c|c|}
\hline
\textbf{Experiment} & \boldmath{$Q^2$ (GeV$^2$)} & \boldmath{$F_\pi(Q^2)$} \\
\hline
JLab (Jefferson Lab) & 1.6 & 0.243+0.012 \\
JLab (Jefferson Lab) & 2.45 & 0.167+0.010 \\
\hline
\end{tabular}
\caption{Experimental values of the pion form factor $F_{\pi} (Q^2)$ at different values of momentum transfer squared $Q^2$ from \cite{Horn}. }
 \label{datatable}
\end{table}

\begin{figure}[htbp]
    \centering
    \includegraphics[scale=0.65]{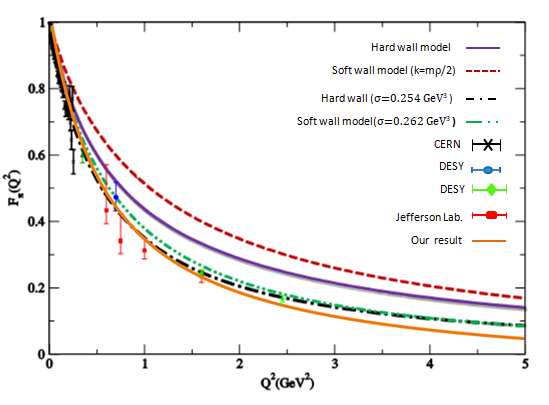}
   \caption{The behavior of the pion form factor $F_\pi(Q^2)$ as a function of momentum squared $Q^2$ (figure adapted from \cite{Herry}). The solid curve corresponds to the prediction from the standard hard-wall model. The dashed curve illustrates the outcome of the conventional soft-wall model with $k = m_\rho/2$. The dash-dotted line depicts the hard-wall scenario with $\sigma = (0.254~\text{GeV})^3$, while the dash-double-dotted line shows the soft-wall framework with $\sigma = (0.262~\text{GeV})^3$. Cross symbols represent a dataset aggregated by CERN \cite{Amendolia}; open circles denote data from DESY, reinterpreted by Ref. \cite{Blok}; the triangles refer to another DESY measurement \cite{Ackermann}; and the  box \cite{Blok} and diamond markers \cite{Horn} correspond to results from the Jefferson Lab. Earlier measurements within the interval $3$--$10~\text{GeV}^2$~\cite{Bebek} are not displayed due to significant uncertainties.
The orange curve is the one we obtain using the thermal soft-wall model at $T=0$ $GeV$.}
\label{13a}
\end{figure}

In Fig. \ref{13a}, we show a comparison of the dependence of the pion EM form factor on $Q^2$, when $T = 0$. We show how our result based on the soft-wall model compares to that obtained by other models and experiments. The solid curve (in purple) represents the standard AdS/QCD hard-wall result, while the dashed line (in red) corresponds to the standard soft-wall model with $k = m_\rho/2$. The dash-dotted (in black) and dash-double-dotted (in green) lines account for the hard-wall and soft-wall models with specific condensate values of $\sigma = (0.254~\text{GeV})^3$ and $\sigma = (0.262~\text{GeV})^3$, respectively. The orange curve illustrates our result using the thermal soft-wall model at $T=0$ $GeV$. What we observer from this comparison is that the thermal soft-wall model produces a normalized form factor profile which is closer to the experimentally observed values, than other theoretical models, at low to intermediate values of $Q^2$. 

\FloatBarrier

\subsection{Pion Thermal Mass}

In Fig. \ref{4a}, we show the temperature dependence of the pion mass for quark flavors $N_{f} =$ 2, 3 and 4.   As mentioned in \autoref{c}, this follows from extrapolating the GMOR relation to finite temperatures (eq. (\ref{gmor})). Thus, we get
\begin{eqnarray}
   M_{\pi} (0) = 0.134 \; GeV
\end{eqnarray}
for the pion mass at zero temperature. Here $m_u = 0.0021$ GeV and $m_d = 0.0046$ GeV are used for the masses of the up and down quark respectively \cite{flag21} (one may also consider a thermal dependence for the quark masses themselves as is done in \cite{pscmass}, although that will not significantly alter our results here); and $\Sigma (0) = - (0.273 \; GeV)^3$ is used for the value of the chiral condensate at $T = 0$ \cite{flag21}.  Note that experimental estimates of $\Sigma (0)$ vary within the range of $- (0.248 \; GeV)^3$ to $- (0.318 \; GeV)^3$ \cite{flag21}, whereas lattice QCD estimates have reported $- (0.273 \; GeV)^3$  \cite{ccvalue}.   
The most widely accepted and experimentally confirmed value for the pion mass is $M_\pi \approx 0.139$ GeV, as reported by the Particle Data Group (PDG) \cite{PDG}, based on experiments such as electron-positron collisions and hadronic interactions. Lattice QCD estimates have reported $M_\pi = 0.140$ GeV \cite{Nikhil}. The GMOR relation thus provides a good leading order estimate of the pion mass at $T = 0$, which approaches the experimental mass for specific values of $\Sigma (0)$.  


For small temperatures, we find that the pion mass is fairly stable and only starts to decrease close to the critical temperature $T_c$ (which itself is flavor dependent). This behavior corroborates other estimates of thermal pions found in Refs.~\cite{Oset, Fern, Xuanmin}.

\begin{figure}[htbp]
    \centering
    \includegraphics[scale=0.5]{6.png}
    \caption{ Temperature dependence of the pion mass based on the thermal soft-wall model, for different quark flavors. }
    \label{4a}
\end{figure}
\FloatBarrier

\begin{figure}[htbp]
    \centering
    \includegraphics[scale=0.8]{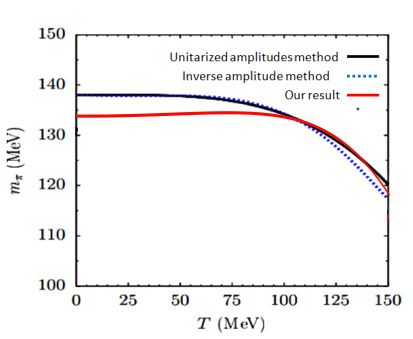}
 \caption{Thermal pion mass obtained from the one-loop pion self-energy correction at
finite temperature \cite{Schenk} using unitarized scattering amplitudes from ChPT (figure adapted from \cite{Juan}). The solid line (black) employs unitarized amplitudes from Reference \cite{Oset}, while the dotted line (blue) uses the Inverse Amplitude Method (IAM) results from Reference \cite{Fern}. The red curve represents our result for the pion mass in the thermal soft-wall model ($T_{c}=0.254$ $GeV$).}
\label{14a}
\end{figure}

More specifically, in Fig. \ref{14a}, we provide a comparison of our result with the thermal pion mass obtained from the one-loop pion self-energy correction at finite temperature using unitarized scattering amplitudes derived from Chiral Perturbation Theory (ChPT) \cite{Juan, Schenk}. The solid line corresponds to the results based on the unitarized amplitudes from Ref. \cite{Oset}, while the dotted line represents the predictions from the Inverse Amplitude Method (IAM) as developed in Ref. \cite{Fern}. For comparison, our result (the red curve) illustrates the pion mass as predicted by the thermal soft-wall model at $N_f = 4$, and fits well with the predictions of  \cite{Juan}.

Note that the pion mass we discuss here is the pole mass, which vanishes at $T_c$ and is ill-defined thereafter. On the other hand, the screening mass continues to grow and remains well-defined beyond $T_c$. The latter has been computed for pions in \cite{Xuanmin} using a hydrodynamic description of the Goldstone degrees of freedom, and also in \cite{pscmass} using the Nambu–Jona-Lasinio model. 

\FloatBarrier

\subsection{Pion Thermal Charge Radius}


In Fig. \ref{4b}, we show the temperature dependence of the pion charge radius within the thermal soft-wall model for quark flavors $N_{f}$ being 2, 3 and 4. This is computed by substituting the temperature-dependent pion EM form factor $F (Q^2, T)$ (eq. (\ref{FFF})) into eq. (\ref{raio}) for the charge radius of the pion, and converting the result into units of fm, where $1 \, GeV^{-1} = 0.197 \, fm$. At $T = 0$, we obtain $r_\pi = 0.73~\text{fm}$ (with the pion decay constant $F = 0.087 \, GeV$ used in \cite{Gutsche2}), and $r_\pi = 0.68~\text{fm}$ (with the pion decay constant $F = 0.093 \, GeV$  used in ChPT following QFT conventions \cite{gasser2010F}). At small temperatures above zero, the charge radius remains fairly stable, and only starts to increase rapidly upon approaching the critical temperature $T_c$, where it diverges - indicating de-hadronization into a quark-gluon plasma. This profile obtained from the thermal soft-wall model is exactly what is also predicted by the thermal sigma model to leading order (one-loop), and also a thermal QCD-Finite Energy Sum Rule \cite{pncoup}, thus validating the thermal holographic QCD approach.

Note that the experimental value of the pion charge radius, as reported by the Particle Data Group (PDG), is given as
$\langle r_\pi^2 \rangle = 0.434~\text{fm}^2 $  $(r_\pi=0.659~\text{fm})$ \cite{PDG}.
This value serves as a widely accepted benchmark obtained from experimental data and phenomenological fits. 
Another study involving pion-electron elastic scattering yields $r_\pi = 0.640(7)~\text{fm}$ \cite{Zhu}. 
In comparison, a recent lattice QCD calculation reports $\langle r_\pi^2 \rangle = 0.42(2)~\text{fm}^2$, which gives $r_\pi=0.648~\text{fm}$ \cite{latticeradiusmass}
which agrees well with the PDG value.

Compared to the experimental estimates above, in the holographic QCD hard-wall model at zero temperature  \cite{Radyushkin2007},  the pion charge radius has been estimated at approximately $r_{\pi}=0.58$ fm, somewhat lower than experimental and lattice estimates. Other authors have reported a value of $0.645$ fm for the hard-wall model after parameter tuning \cite{Herry}. On the other hand, investigations into the (standard) AdS soft-wall model at zero temperature have reported the pion charge  radius to be  $0.494~\text{fm}$, as well as $0.60$ fm after parameter tuning \cite{Herry}. Although this is close to the radius estimated by the hard-wall model at $0.58~\text{fm}$ \cite{Radyushkin2007}, it is still lower than the experimental value of $0.659~\text{fm}$ \cite{PDG}. In fact, it has been pointed out recently in \cite{Martn} that  standard formulations of both, the hard-wall and the soft-wall models underestimate the pion charge radius. To rectify this, a deformed AdS soft-wall model was suggested, in which the dilaton slope is specific to each field in the bulk. That is, besides a dilation slope associated to the pion $\kappa_{\pi}$, an additional dilation slope is introduced for the photon $\kappa_{\gamma}$ \cite{Martn}. This allows for an additional free parameter that helps reduce the discrepancy of the pion charge radius compared to the experimental value. With this the authors of \cite{Martn} report a  $2 \, \%$ relative error (after introducing a momentum dependence in their $\kappa_{\gamma}$). However, this comes at the cost of introducing an additional parameter in the model, as well as modifying the form factor post hoc. Alternatively,  other authors have suggested using conformal dimension \( \Delta = 2 \) (instead of 3) to fix the pion charge radius based on the light-front soft-wall model \cite{Brodsky2008}. 

In our approach here, we do not introduce additional parameters into the model. We use the expression for the pion's thermal form factor in eq. (\ref{FFF}). This is derived from the overlap integral between the thermal scalar mode, dual to the pion, and the vector mode, dual to the photon (following the zero-temperature formulation in \cite{Erlich, Karch, Martn}). We then find $r_\pi = 0.73~\text{fm}$ at $T = 0$ (when $F = 0.087 \, GeV$), and $r_\pi = 0.68~\text{fm}$ (when $F = 0.093 \, GeV$). These values are $10\%$ and $3\%$ away, respectively, from the experimental value. On the other hand, if one chooses to tune the pion decay constant in the soft-wall model to $F =  0.097 \, GeV$, then that would exactly match the experimental value of the pion charge radius!

\begin{figure}[htbp]
    \centering
\includegraphics[scale=0.5]{8.png}
    \caption{  Temperature dependence of the pion charge radius based on   the thermal soft-wall model, for different quark flavors. }  
    \label{4b}
\end{figure}

\FloatBarrier

\subsection{Pion Generalized Parton Distribution}

The GPD of the pion has been primarily determined on the basis of theoretical models and simulations. These measurements are not directly obtained through experiments, but are estimated using lattice QCD and constructive experimental models.
%
%
Further experiments (e.g., JLab and COMPASS) will be able to provide more precise measurements of the pion's GPD by comparing them with theoretical predictions such as the ones we present in this subsection.

\begin{figure}[htbp]
    \centering
    \includegraphics[scale=0.5]{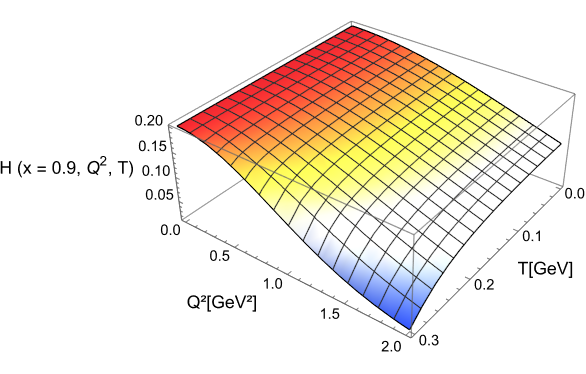} \hspace{1em}
    \includegraphics[scale=0.5]{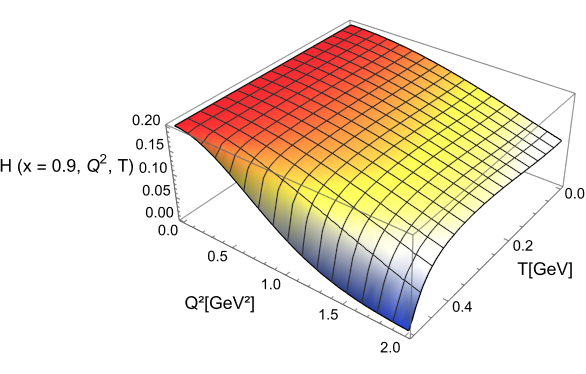}
    \par\textbf{$(a)$  $\alpha=0.2$} \hspace{12em} \textbf{$(b)$ $\alpha=0.4$}\par\vspace{1em}
     \caption{ Temperature  and momentum transfer squared dependence of the pion's GPD for   $\alpha = 0.2 $ \;  $(a)$ \text{and} \; $\alpha = 0.4 $ $(b)$ respectively, at fixed value of $x = 0.9$. } 
     \label{16alp}
\end{figure}

We will consider $N_f = 2$ in this subsection, and $\alpha = 0$ until  we discuss Fig. \ref{16alp}.  
In Fig. \ref{7a} we show the $Q^{2}$- and $T$-dependence of the pion GPD at $x = 0.1, \, 0.5, \, \text{and} \;  0.9$ respectively.  
These results highlight the intricate relationship between the Bjorken scaling variable \( x \), momentum transfer squared \( Q^2 \), and temperature $T$. Note that here we only consider the unpolarized pion GPD, since we do not need to take into account the spin of the particle.

\begin{figure}[htbp]
    \centering
    \includegraphics[scale=0.40]{9a.png} \hspace{1em}
    \includegraphics[scale=0.40]{9b.png}
    \par\textbf{$(a)$ $x=0.1$} \hspace{20em} \textbf{$(b)$ $x=0.5$}\par\vspace{4em}

    \includegraphics[scale=0.45]{9c.png}
    \par\textbf{$(c)$ $x=0.9$}\par\vspace{0.5em}

    \caption{Temperature and momentum squared dependence of the pion GPD for different values of $x$. }
      \label{7a}
\end{figure}

\FloatBarrier

\begin{figure}[htbp]
    \centering
  \includegraphics[scale=0.4]{10a.png}
   \hspace{1em}
  \includegraphics[scale=0.6]{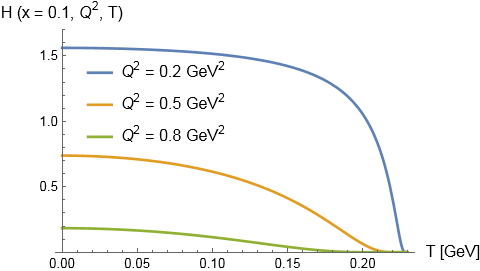}
    \par\textbf{$(a)$ $x = 0.1$} \hspace{12em} \textbf{$(b)$  $x = 0.1$ }\par\vspace{1em}
       \caption{(a): $Q^{2}$ dependence of the pion GPD at  different value of $T$  and  (b): $T$ dependence of the pion GPD at different $Q^2$. }
    \label{gpd11}
\end{figure}

Fig. \ref{gpd11} shows 2D slices of the profiles seen in Fig. \ref{7a} above. We first show the $Q^{2}$ dependence of the GPD at  $T=0$ $GeV$, $0.15$ $GeV$, and $0.2$ $GeV$  taking a fixed $x = 0.1$.  We then show the temperature dependence of the GPD for $Q^{2}= 0.2$ $GeV^{2}$, $0.5 $ $GeV^{2}$, and $0.8$ $GeV^{2}$ with $x = 0.1$.

\FloatBarrier

Taken together, these plots indicate that at high temperatures, the decrease in GPD values, especially at large momentum transfer \( Q^2 \) reflects weaker quark binding, reduced coherence in parton distributions, and partial deconfinement.  The temperature and momentum dependence of the pion's GDP provides insight into the internal structure of the pion at varying temperatures.

In Fig. \ref{8a}, we illustrate the $T$ and $x$ dependence of the pion GPD  at different momenta squared $ Q^2=0.5 \; GeV^2$, $ Q^2=1 \; GeV^2$, and $ Q^2=1.5 \; GeV^2$.
This provides information on how finite temperature affects the internal structure (quarks and gluons) of the pion. Various values of \(Q^2\) reflect changes in the pion's structure depending on the scale and thermal conditions. As the temperature increases, an increase in \(x\) results in a decrease in the value of the pion's GPD, which confirms the influence of thermal and dynamic effects on the pion's structure.

\begin{figure}[htbp]
    \centering
    \includegraphics[scale=0.41]{11a.png} \hspace{1em}
    \includegraphics[scale=0.41]{11b.png}
    \par\textbf{$(a)$  $Q^{2} = 0.5 \; GeV^{2}$} \hspace{8em} \textbf{$(b)$ $Q^{2} = 1 \; GeV^{2}$}\par\vspace{1em}

    \includegraphics[scale=0.45]{11c.png}
    \par\textbf{$(c)$ $Q^{2} = 1.5 \;   GeV^{2}$}\par\vspace{0.5em}

    \caption{ Temperature and $x$ dependence of the pion GPD  at  different momentum transfer  $Q^2$. }
    \label{8a}
\end{figure}

\FloatBarrier

\begin{figure}[htbp]
    \centering
  \includegraphics[scale=0.55]{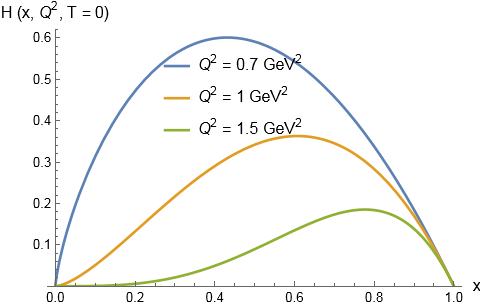}
   \hspace{1em}
  \includegraphics[scale=0.4]{12b.png}
    \par\textbf{$(a)$ $T = 0$ $GeV$} \hspace{12em} \textbf{$(b)$  $Q^{2} = 1$ $GeV^2$ }\par\vspace{1em}
       \caption{(a): The $x$ dependence of the GPD at  $T=0$ $GeV$ for a range of $Q^{2}$ values. (b): The  $T$ dependence of the GPD at fixed $Q^{2}= 1$ $GeV^{2}$ for a range of $x$ values.}
    \label{9a}
    \label{H11}
\end{figure}

Then, in Fig. \ref{9a}, the dependence of the GPD on $x$ is examined at  $T=0$ $GeV$ for varying momentum transfers  
$Q^{2}=0.7$ $GeV^{2}$, $1$ $GeV^{2}$, and $1.5$ $GeV^{2}$. Our results in this figure compare nicely to the zero temperature GPD discussed in Ref. \cite{Nav}, which was also studied in the context of the AdS soft-wall model. Here we have used $\zeta = 0$ for the skewness parameter, whereas the authors of \cite{Nav}  consider $\zeta \neq 0$.   
Additionally, the temperature dependence of the pion  GPD is analyzed at a fixed momentum transfer squared of $Q^2 = 1~\mathrm{GeV}^2$ for selected values of the Bjorken variable $x = 0.2, 0.4, 0.6$. 

\FloatBarrier

In Fig. \ref{6a}, we analyze the $Q^2$- and $x$-dependence of the pion GPD at temperatures $T=0$ $GeV$ versus $T=0.2$ $GeV$ respectively.

\begin{figure}[htbp]
    \centering
   \includegraphics[scale=0.38]{13a.png} \hspace{1em}
    \includegraphics[scale=0.38]{13b.png}
    \par\textbf{$(a)$  $T=0$ $GeV$} \hspace{12em} \textbf{$(b)$ $T=0.2$ $GeV$}\par\vspace{1em}
      \caption{$Q^2$ and $x$ dependence of the pion GPD at different temperatures.}
    \label{6a}
\end{figure}

\FloatBarrier

\begin{figure}[htbp]
    \centering
  \includegraphics[scale=0.6]{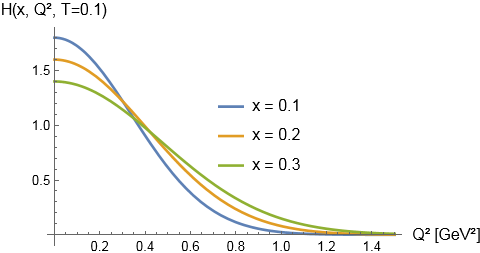}
   \hspace{1em}
    \includegraphics[scale=0.6]{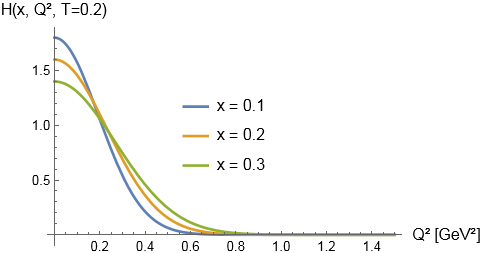}
    \par\textbf{$(a)$ $T = 0.1$  $GeV$ } \hspace{12em} \textbf{$(b)$  $T = 0.2$  $GeV$ }\par\vspace{1em}
        \caption{   The $Q^{2}$ dependence of the GPD at  different values of $x$ when (a) $T = 0.1 \, GeV$ and (b) $T = 0.2 \, GeV$.  }
    \label{10a}
\end{figure}

Fig.~\ref{10a}, showcases 2D sections of the profiles shown in Fig.~\ref{6a} above. We depict the $Q^2$ behavior of the pion GPD at fixed \( x = 0.1, \, 0.2, \, 0.3 \) for temperatures, \( T = 0.1 \, \mathrm{GeV} \) and \( 0.2 \, \mathrm{GeV} \) respectively.

At \( Q^2 = 0\, GeV^2\), the GPD values are highest for small \( x \), following the hierarchy:
\[
H(x=0.1)=1.8 > H(x=0.2)=1.6 > H(x=0.3)=1.4,
\]
which holds for both temperatures. This reflects enhanced parton densities at small momentum fractions.

As \( Q^2 \) increases, the GPD decreases and approaches values consistent with thermal suppression. At \( T = 0.1\, \mathrm{GeV} \), this suppression occurs earliest for \( x = 0.1 \), followed by \( x = 0.2 \) and \( x = 0.3 \). The same trend persists at \( T = 0.2\, \mathrm{GeV} \), but the onset of suppression occurs at lower \( Q^2 \), indicating stronger thermal effects at higher temperature.

\FloatBarrier

These results show that small-\( x \) modes are more sensitive to the hot medium, and that increasing temperatures lower the resolution scale at which the pion’s parton structure starts to exhibit critical behavior. This may be interpreted as a signal of partial deconfinement or medium-induced modifications to hadronic matter.  These findings highlight the role of thermal effects in modifying the pion's GPD, particularly in the small-\( x \) and low-\( Q^2 \) regime, where signatures of early deconfinement may emerge.

Finally, in Fig. \ref{16alp}, we show the temperature and momentum dependence of the pion's GPD for different $\alpha = 0.2 \; \text{and} \; 0.4$ respectively, at fixed $x = 0.9$. The parameter $\alpha$ captures  gravitational effects by scaling the prefactor in the action that encodes the position of the event horizon in the AdS-Schwarzschild geometry.    
In Fig. \ref{16alp}, we see that $\alpha$ increases the pion's GPD at fixed $x$, $Q^2$ and $T$, which ultimately leads to an increase in the form factor (as seen earlier in Fig. \ref{2a}).  $\alpha$ also increases the critical temperature $T_c$, which can be seen by the wider range of $T$ in Fig. \ref{16alp}, after which the GPD tapers off. Hence, $\alpha$ acts against the de-hadronization of pions within a hot medium. This corroborates what is expected when gravitational back-reaction upon a gas of hadrons in an AdS geometry leads to a transition of the gas to denser phases of matter, such as in a neutron star or black hole \cite{deBoer, XDA1}.
   
\FloatBarrier

\subsection{Pion Electric Charge Density}

Next, we examine the charge density distribution of pions at different temperatures. This provides crucial insight into their internal structure and the spatial distribution of quark constituents. The $\pi^+$ meson is composed of a $u$ and $\bar{d}$ quark, while the $\pi^-$ meson consists of a $\bar{u}$ and $d$ quark. Due to their opposite electric charge, the spatial distribution of their charge densities is expected to be symmetric but sign-reversed.  

More specifically, the charge density distribution $\rho(b, T)$ of the positive pion ($\pi^+$) is shown in Fig. \ref{7a} as a function of the radial distance and temperature. As the temperature increases, the absolute value of the charge density decreases, and the peak of the distribution becomes less pronounced.   
This behavior can be interpreted in the context of the confinement--deconfinement transition in QCD. 
At low temperatures, quarks are confined within hadrons, resulting in a well-localized and sharply peaked charge density. 
As the temperature increases and the system approaches the deconfinement regime, the pion gradually loses its compact structure, leading to a broader and smoother spatial distribution of its charge density.  In other words, the observed decrease in the magnitude of the charge density and the flattening of its peaks with increasing temperatures indicate a gradual transition from the confined hadron phase to the deconfined quark-gluon plasma phase. 

\begin{figure}[htbp]
    \centering
    \includegraphics[scale=0.6]{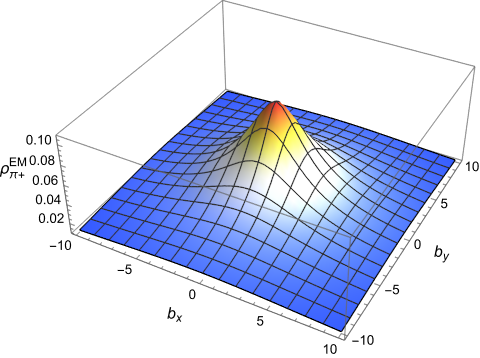} \hspace{1em}
    \includegraphics[scale=0.6]{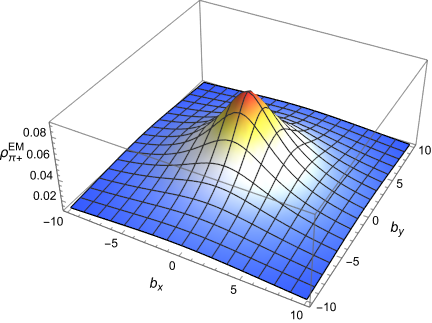}
    \par{$(a)$ $T=0$ $GeV$ } \hspace{15em} {$(b)$ $T=0.15$ $GeV$ }\par\vspace{3em}
    \includegraphics[scale=0.6]{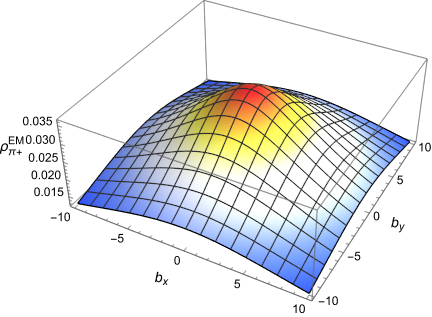}
    \par{$(c)$ $T=0.22$ $GeV$ }\par\vspace{0.5em}

    \caption{  The spatial distribution of the $\pi^+$ meson electric charge density, based on the parameters $b_x$, $b_y$, at different values of $T$. } 
      \label{7a}
\end{figure}

\FloatBarrier

\begin{figure}[htbp]
    \centering
 
 \includegraphics[scale=0.55]{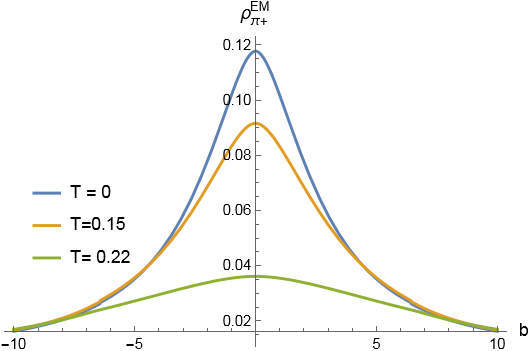}
\hspace{1em}
 \includegraphics[scale=0.55]{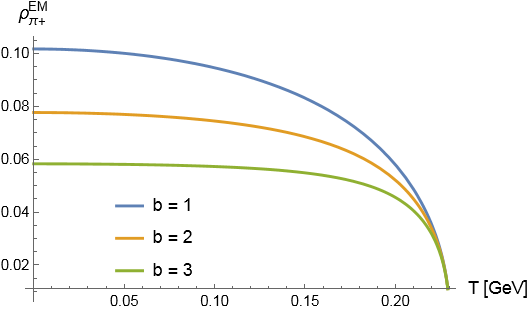}

    \par\textbf{$(a)$ } \hspace{20em} \textbf{$(b)$  }\par\vspace{1em}
       \caption{  (a) 2D illustration of the $\pi^+$ meson charge density at different $T$; and (b) the temperature dependence of pion charge density at different values of $b$. } 
      \label{12aa}
\end{figure}

In Fig.~\ref{12aa} we examine 2D projections of the 3D charge density distributions seen above. Panel (a) of Fig.~\ref{12aa}  shows the dependence of the pion charge density on the parameter $b$ for several temperatures $T$. Panel (b) illustrates the temperature dependence of the charge density for different values of $b$. We find that as the temperature increases, the charge density distribution exhibits a clear spatial delocalization. Close to the critical temperature, the pion charge becomes diffuse, indicating de-hadronization.



\FloatBarrier

\begin{figure}[htbp]
    \centering
 
 \includegraphics[scale=0.55]{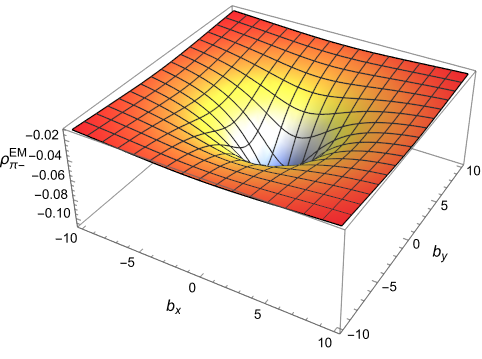}
\hspace{1em}
 \includegraphics[scale=0.55]{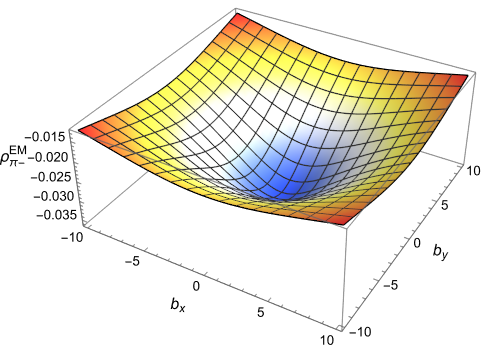}

    \par\textbf{$a)$ $T=0.1$ $GeV$ } \hspace{12em} \textbf{$b)$  $T=0.22$ $ GeV$  }\par\vspace{1em}
       \caption{ Charge density of $\pi^{-}$ pion  at different temperatures. }
      \label{12a}
\end{figure}

In Fig \ref{12a}, the dependence of the negative pion ($\pi^-$) charge density on both the parameter $b$ and temperature $T$ has been analyzed. 
The behavior of the $\pi^-$ charge density exhibits a symmetry with respect to the corresponding dependence of the positive pion ($\pi^+$). 
Specifically, the variations of the $\pi^-$ charge density with $b$ and $T$ mirror those of the $\pi^+$, 
indicating that the spatial distribution and thermal response of the negative pion are qualitatively similar to the positive pion, 
but with opposite sign in charge. 
This symmetry confirms the expected particle-antiparticle behavior and is consistent with the underlying QCD confinement dynamics.

\FloatBarrier

\subsection{Pion-Baryon Coupling Constants}

\begin{figure}[htbp]
    \centering
    \includegraphics[scale=0.55]{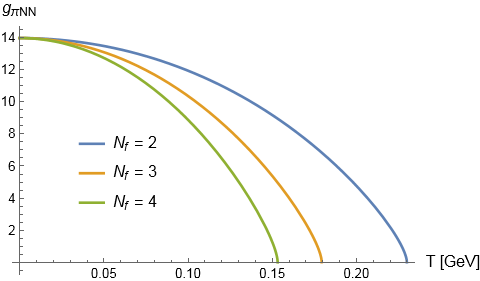}
    \hspace{1em}
    \includegraphics[scale=0.55]{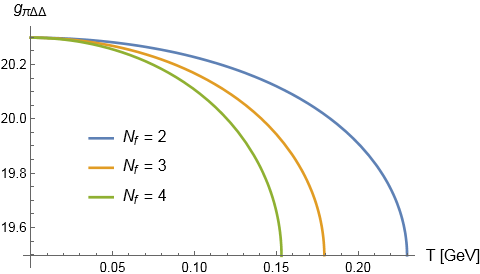}
    
    \par\text{$(a)$  $g_{\pi NN}$ coupling constant} \hspace{8em} \text{$(b)$ $g_{\pi \Delta \Delta}$ coupling constant}\par\vspace{1em}

    \caption{$T$ dependence of (a) the pion-nucleon, and (b) the pion-$\Delta$ baryon coupling constants. }
    \label{5a}
\end{figure}

In Fig.~\ref{5a}, we explore the temperature dependence of the pion-nucleon \( g_{\pi N N}(T) \), and pion-$\Delta$ baryon \( g_{\pi \Delta \Delta}(T) \) coupling constants. It is observed that the values of these constants decrease with increasing temperature. This result for \( g_{\pi N N}(T) \) is comparable to that obtained in Ref. \cite{pncoup}.
We also study the temperature dependence of these couplings for different numbers of quark flavors \( N_f \) at a fixed value of $F=0.087$ $GeV$  as in Refs.~\cite{Gutsche2, Gutsche3}. We find that the values of these coupling constants increase as \( N_f \) decreases.  
As $N_f$ decreases, reduced fermionic screening enhances the effective gauge coupling at low energies, as captured by the one-loop beta function where gauge self-interactions dominate. Consequently, for $N_f = 2$, the stronger gauge interactions lead to a higher critical temperature for deconfinement and chiral symmetry restoration compared to larger values of $N_f$.


The pion-nucleon coupling constant \( g_{\pi NN} \) is found to be approximately $13.5$ from 
 pion-nucleon scattering  \cite{Ahn}. 
Although the primary objective of our work is to determine how physical quantities related to pions, change with increase in the medium's temperature, we also examine how the results of the model (specifically the standard soft-wall model) at \( T = 0 \) aligns with corresponding values obtained from other models for these constants. 
In different theoretical models, the value of the coupling constant \( g_{\pi NN} \) was found to range approximately from 13 to 14. Specifically, the Yukawa model estimates that for charged pions   \( g_{\pi NN}=14.52(26)\) \cite{Babenko}. Lattice QCD simulations provide an estimate of \( g_{\pi NN} = 13.8 \pm 5.8 \) \cite{Lattice}, and \( g_{\pi NN} = 13.7  \) in Ref. \cite{Bugg}. On the other hand, models such as the hard-wall model have free parameters $\kappa_1$, $\kappa_2$, $\kappa_3$ and $\kappa_4$ in their expressions for the zero-temperature coupling constants (between pions and baryons). 
In the hard-wall model, these parameters are fixed to reproduce values of the coupling constants close to the experimental results. Following the same procedure in our model, we obtained suitable values for the couplings as well.
 In the thermal soft-wall model, in order to reproduce the empirical value of the coupling constant, we find that $\kappa_1=-5.7$ and $\kappa_3=1.17$,  while $\kappa_2$ and $\kappa_4$ remain unconstrained  since the second term in the Lagrangian in eqs.~(\ref{p43}) and (\ref{53}) vanishes. With these parameter choices, our calculation yields a zero-temperature pion-nucleon coupling constant of $g_{\pi NN} = 13.95$. 

For the \( g_{\pi \Delta \Delta} \) coupling, we have derived a value  $20.3$  which is in good agreement with the value of \( g_{\pi \Delta \Delta} \approx 20 \) reported in Ref. \cite{PDG}.
Note that these values are specific to the type of interaction under consideration.  In all of our numerical analysis, for the  light quark mass $m_{q}$ and quark condensate $\Sigma$, we have used values $m_q=0.00334$ GeV and $\left(\Sigma\right)^{1/3}= 0.273$ $ GeV$ respectively \cite{Maru} with the value of the Yukawa coupling being   $g_{Y}=14.52$.


\FloatBarrier

\section{Discussion}
\label{f}
This work presents a systematic investigation of nonperturbative low energy dynamics of pions in a hot medium, based on the recent thermal soft-wall model of holographic QCD. The thermal soft-wall model improves upon the original soft-wall model by introducing a temperature-dependent dilation field  \cite{Gutsche2, Gutsche3}.  In the dual 4-dimensional gauge theory, this enables the calculation of thermal averages of QCD $n-$point correlation functions. In particular, the $3$-point function involving the pion current is of particular significance as it involves the pion coupling to other baryons during elastic scattering processes. More specifically, we have computed the theoretical expression for the temperature-dependent pion-nucleon-nucleon, as well as, pion-$\Delta$-$\Delta$ coupling. What we find is that these expressions are analogous to the expressions found earlier in the zero-temperature case \cite{Ahn}, but with the modification that all fields are now temperature dependent.  As expected, we find that these couplings decrease with increasing temperature and vanish near the confinement-deconfinement temperature, where hadrons undergo a phase transition to a quark-gluon plasma. These results compare well with other models and experimental observations. 

Furthermore, we have numerically explored temperature profiles of the pion form factor, GPD, mass, charge radius and charge density. The objective in this part was to validate phenomenological features of the thermal soft-wall model compared to other models such as the hard-wall model and lattice QCD, as well as compared to experimental values of these nonperturbative observables. 

In Table~\ref{tab:pion-comparison}, we present a comparison of our results for the pion mass, coupling constant (with nucleons), and pion charge radius at zero temperature versus other theoretical models and experiments.

\begin{table}[htbp]
\centering
\begin{tabular}{|c|c|c|c|}
\hline
\textbf{Model/Method} & \textbf{Mass ($\text{GeV}^{2}$)} & \textbf{Pion-Nucleon Coupling} & \textbf{Charge Radius ($\text{fm}$)} \\
\hline
Thermal Soft-Wall Model & 0.134 & 13.95 & 0.73 ($F = 87~MeV$), 0.68 ($F = 93~MeV$)  \\
\hline
Hard-Wall Model & 0.14 \cite{Maru}& 13.5 \cite{Ahn}, 13.6 \cite{Maru}& 0.58 \cite{Radyushkin2007}, 0.645 \cite{Herry}  \\

\hline
Lattice QCD & 0.14 \cite{latticeradiusmass} & $13.8 \pm 5.8$ \cite{Lattice}  & 0.648 \cite{latticeradiusmass} \\
\hline
Experiment & 0.13957 \cite{PDG} & $13.4 - 13.5$ \cite{PDG} &  0.659 \cite{PDG}\\
\hline
\end{tabular}
\caption{Comparison of the pion mass, coupling constant and radius at $T=0$ $GeV$, obtained using the thermal soft-wall model, versus other theoretical models and experimental data.}
\label{tab:pion-comparison}
\end{table}

Moreover, as shown in Fig.~\ref{13a}, the momentum squared dependence of our  normalized pion form factor corroborates with other models and  shows good agreement with experimental values, though the exact matches may depend on the specific normalization and parameters chosen for the model (as detailed in the numerical results). This comparison highlights the thermal soft-wall model's consistency and predictive power in capturing pion low-energy dynamics.
Besides the $T = 0$ values mentioned above, the temperature profiles (in Section 5 above) of the pion GPD, form factor, mass, radius and charge density, obtained using the thermal soft-wall model, are consistent with what one expects at increasing temperature: that is, they gradually fall off at higher temperatures (except for the radius, which diverges as explained in Section 5).  In conclusion, our results present a comprehensive exploration of the temperature and momentum dependence of low-energy observables of pion dynamics, contributing to the ongoing investigation of hadron interactions within a hot medium, while validating the phenomenological applicability of the thermal soft-wall model. 

We note that suitable extensions to the formalism of the thermal soft-wall model may also be considered, either by admitting higher-order terms in the temperature dependence of the dilaton field through $K^2 (T)$ and / or considering a more general form of the dilation field. That would increase the precision of the model at higher temperatures, including in the vicinity of the chiral phase transition. Another interesting extension has been considered in \cite{Chelabi1, Chelabi2}  where the bulk scalar potential was  replaced with a quartic (and cubic) one. This provides a natural mechanism for spontaneous symmetry breaking, using which, a non-vanishing chiral condensate was computed near the chiral limit. Of course, this comes at the cost of introducing two additional free parameters, whose phenomenological consequences were studied in \cite{Xuanmin}.  While these extra parameters may provide greater flexibility with model-fitting, it would be interesting to investigate their microscopic origins. 

An interesting generalization of the AdS/QCD thermal soft-wall model would be to consider a holographic neutron star in AdS space, rather than a black hole. Such holographic composites have been studied in the general context of AdS/CFT  \cite{deBoer, XDA1, NG1, NG2}. The holographic dual of such a star in the bulk would be a strongly interacting finite-temperature quantum fluid on the boundary, where the boundary theory corresponds to finite-temperature QCD, rather than a strictly conformal field theory. Furthermore, gravitational collapse of such a neutron star to a black hole, suggests the holographic dual of the mechanism leading to chiral symmetry breaking via the confinement-deconfinement phase transition on the boundary QCD. In a similar vein, one may investigate the AdS/QCD dual of black rings in Taub-NUT-AdS \cite{XDA2, XDA4}, or fragmented AdS geometries \cite{XDA3, XDA4} (both of which are asymptotically AdS), which would presumably refer to highly entangled topological phases of QCD matter.

\section*{Acknowledgments}
 \textit{We dedicate this work to the fond memory of Prof. Ivan Schmidt, who has been a dear mentor and an inspiration to us. } 
 
 N.Nasibova would like to express sincere gratitude to Professor Francesco Giacosa at the Institute of Physics, Jan Kochanowski University,  Kielce, Poland, for the hospitality, where a part of this research was carried out. We also thank Professor Valery Lyubovitskij from the Institut für Theoretische Physik, Universität Tübingen, Germany; Professor Tahmasib Aliyev from Middle East University, Türkiye; and Professor Shakir Nagiyev from the Institute of Physics of the Ministry of Science and Education, Azerbaijan; for useful suggestions and feedback.

\end{document}